\documentclass[]{jfm}

\usepackage{graphicx}
\usepackage{newtxtext}
\usepackage{newtxmath}
\usepackage{natbib}
\usepackage{hyperref}
\hypersetup{
    colorlinks = true,
    urlcolor   = blue,
    citecolor  = black,
}

\newcommand{\RomanNumeralCaps}[1]
\linenumbers

\title{Uncovering triadic interaction relationships latent in Mode A for the wake of a circular cylinder}

\author{Yuto Nakamura
  \corresp{\email{yuto.nakamura.t4@dc.tohoku.ac.jp}},
  Shintaro Sato,
 \and Naofumi Ohnishi}

\affiliation{\aff{1}Department of Aerospace Engineering, Tohoku University,
Sendai, 980-8579, Japan}

\begin{document}
\maketitle

\begin{abstract}
Coherence between multiple low-frequency components latent in the flow fields characterizes the nonlinear aspects of fluid dynamics. This study reveals the existance of the distinct frequency components and their interaction relation of the classical Mode A of the cylinder wake. Primaries are one-third of the Karman vortex shedding frequency (third-subharmonic) and bubble pumping, known as the previous study. However, when the spanwise domain size in numerical simulations is sufficiently large, their interaction is obscured by the presence of numerous frequency components. To address this, we introduce a process in which distinct frequency components gradually emerge by starting with a small spanwise domain size and then gradually increasing it from $3.3D$ to $4.7D$, where $D$ represents the diameter of the cylinder. From $3.3D$ to $3.5D$, only the vortex shedding frequency harmonics are present. Third-subharmonic frequency appeared ranging from $3.5D$ to $3.7D$. Bispectral mode decomposition reveals that the harmonics of the third-subharmonic frequency govern the flow in this domain size. The bubble pumping is emergence in the flow fields between $3.7D$ and $3.8D$. The frequency component after this emergence is not only the harmonics of bubble pumping and periodic nature is disrupt. Nonlinear interactions between bubble pumping, the Karman vortex, and the third-subharmonic component complicate the temporal behavior of the flow field. Utilizing the constraint of the spanwise domain size, our approach effectively reveals the interaction relationship between frequency components inherent a flow field with a significant number of frequency components.

\end{abstract}

\begin{keywords}
%Authors should not enter keywords on the manuscript, as these must be chosen by the author during the online submission process and will then be added during the typesetting process (see \href{https://www.cambridge.org/core/journals/journal-of-fluid-mechanics/information/list-of-keywords}{Keyword PDF} for the full list).  Other classifications will be added at the same time.
\end{keywords}

%{\bf MSC Codes }  {\it(Optional)} Please enter your MSC Codes here

\section{Introduction}\label{intro}
The fluctuation in the flow fields at relativity low frequencies compared to the dominant frequency component has been observed in a various flow situation \citep{Chang_1994, Zelman_1993, Yokota_2025}. The low-frequency component exhibits a large-scale spatial structure and influences a wide range of frequencies across the flow field. Therefore, identifying its presence and effect on other frequency components is crucial for a comprehensive understanding of the flow dynamics. In the flow fields around an object, low-frequency fluctuations of about one-tenth of the Karman vortex shedding frequency $St$ have been discovered from the time series data of flow fields, drag, and lift coefficients \citep{NAJJAR_1998,Jiang_2017,cylinder4}. This low-frequency fluctuation is frequently observed on complex flow fields, such as high Reynolds number ($Re$) cases. In recent years, the low-frequency component has become detectable through experiments and numerical simulations with the aid of data-driven science \citep{Yokota_2024,ohmichi2019numerical,Okano_2024}. However, the conditions under which low-frequency components appear and their effect on the flow field are not fully understood. 

The low-frequency component was first detected by the wind tunnel experiment reported by \citet{roshko1954development}. In his experiment, the velocity fluctuation in the flow fields behind a circular cylinder was measured at $Re$ values ranging from $40$ to $10,000$. After $Re=150$, it is observed that the flow fields transition to the three-dimensional flow, whose Karman vortex is non-uniformly distributed in a spanwise direction. With the transition to three-dimensional flow, irregular bursts at low frequencies were reported in the time variation of velocity fluctuation.  These irregular bursts became more frequent as $Re$ increased, and they persisted until $Re=300$. When $Re$ exceeds $300$, the velocity fluctuation becomes completely irregular, and flow fields become turbulent. Therefore, the flow fields at $150 \leq Re \leq 300$ was classified as a transition region. Based on this experiment, the emergence of low-frequency components can be marked as the beginning of the turbulence. 

After the experiment by \citet{roshko1954development}, flow fields behind a circular cylinder and its low-frequency behavior have been widely studied. \citet{Williamson_1989} demonstrated that the critical $Re$ of the transition to three-dimensional flow was dependent on whether the wake vortex shedding was parallel or oblique to the cylinder. In the case of parallel vortex shedding, the transition region can be divided into two stages, and the flow that forms at a $Re$ around $190$ is known as Mode A \citep{williamson1988existence}. Contrary to this, the experiment of \citet{roshko1954development} was classified as a case of oblique vortex shedding due to the effect of the spanwise end of the cylinder model in the wind tunnel. It was also pointed out that the low-frequency component observed in the experiment may have originated from the spanwise end of the model. 

From the numerical aspects, numerous studies have computed the flow field in the transition region. \citet{Karniadakis_1992} showed that the transition to a three-dimensional flow was owing to the secondary instability derived from the two-dimensional Karman vortex. Based on their result, \citet{Noack_1993} introduced the Floquet theory to determine the stability of periodic flow. Following the Floquet theory, \citet{Barkley_1996} investigated the spanwise wavelength stability of the Karman vortex using a Floquet analysis \citep{Noack_1993}. The stable spanwise wavelength obtained from the Floquet analysis was consistent with the previous experimental results \citep{Williamson_modeA,wu1994experimental}. 

On the side of uncovering low-frequency behavior, the existence of low-frequency fluctuation is identified in the vortex of stable spanwise wavelengths based on the numerical investigation of \citet{Henderson}. The stable spanwise wavelengths in his numerical investigation assume a periodic to the spanwise direction. Thus, this result shows that the low-frequency component appears regardless of the spanwise end of the cylinder in the experiment. Henderson denotes that the low-frequency fluctuation is not the presence of a low-frequency component but rather due to multiple frequencies distinct from, yet close to, $St$. This result indicates that low-frequency beating may appear, resulting in the formation of complex flows with multiple frequencies. However, no clear evidence was obtained.

With the advancement of computers, direct numerical simulation (DNS) in transition regions has become feasible \citep{modeA,Jiang_2017,cylinder4}. In the DNS, the effect of the spanwise computational domain size is not negligible since Mode A has a relatively large spanwise wavelength to the cylinder diameter. \citet{Jiang_2017} quantitatively investigated the effect of spanwise domain size and boundary conditions on the Mode A flow properties in the DNS at $Re=200$. They reveal the temporal behavior of Mode A affected by the spanwise domain size $L_z$. For instance, low-frequency fluctuations do not exist with small spanwise domain sizes. They concluded that the low-frequency fluctuation is caused by repetitive shifts in the spanwise wavelength to other wavelength values, and it changes their frequency. The repetitive shifts can be constraint in the small dmain size because the spanwise wavelength is limited by the spanwise domain size.

Based on the insights from previous numerical simulations of \citet{Henderson,Jiang_2017}, the low-frequency fluctuation is associated with the presence of multiple frequencies close to $St$. However, the existence of a spatial structure corresponding to low-frequency fluctuations has not been confirmed in Mode A. In the case of multiple frequencies existing in the flow fields, their frequency components are considered to interact due to the nonlinear terms in the Navier-Stokes equations \citep{phillips_1960,yeung_2024,freeman_2024}. In short, the nonlinear term $\mathcal{N}(f_p,f_q)$ derived from two frequency component $f_p$, $f_q$ is
\begin{equation}
\mathcal{N}(f_p,f_q)=A_{f_p}e^{(f_pt+\theta_p)i}A_{f_q}e^{(f_qt+\theta_q)i}=A_{f_p}A_{f_q}e^{(f_p+f_q)ti+(\theta_p+\theta_q)i},
   \label{intrnon}	
\end{equation}
where $t$ denotes time, $\theta_f$ is temporal phase, $A_f$ is spatial structure of frequency $f$ component. Hence, the nonlinear interaction of two frequency components produces different frequency components $f_p+f_q$. The difference interaction between two frequencies $f_p$, $f_q$ that have close frequencies produces a low-frequency component $f_p-f_q$. 

For analyzing numerical results, the data-driven science \citep{POD1,yeung_2024,POD0,Glazkov_2024} has become a powerful tool due to the development of computers. The dynamic mode decomposition (DMD) \citep{DMD,Tu_2013} is the practical method for extracting the coherence structure of specific frequency, including low-frequency components, from flow data.  In the context of nonlinear interaction, \citet{schmidt_2020} proposed the bispectral mode decomposition (BMD) for detecting the nonlinear interaction relations in the flow data based on bispectrum. The DMD and BMD are reasonable approaches to extracting the coherent structure for low-frequency fluctuation and a nonlinear interaction relationship between the $St$ and multiple-frequency components close to $St$.

This paper investigates the variation in the coherent structure of Mode A due to the appearance of low-frequency fluctuations using the DMD and BMD. As a typical example of Mode A, we consider the flow field around a circular cylinder with a $Re=200$, the same condition as \citet{Jiang_2017}. Based on the observation of \citet{Jiang_2017}, low-frequency fluctuation can be suppressed by the spanwise domain constraints. Thus, this constraint enables us to investigate the effect of the low-frequency component on Mode A properties since Mode A is obtained before and after low-frequency fluctuations existing by gradually changing the spanwise domain size in the numerical simulation. 
This novel approach offers unprecedented insights into the fundamental properties of low-frequency components. Furthermore, the effect of low-frequency components on flow fields can be easily estimated because the spanwise boundary constraints simplify the flow field.

The structure of this paper is as follows. Section \ref{numerical} describes the methodology of numerical simulation and modal analysis of DMD and BMD. Section \ref{results} presents simulation, DMD, and BMD results. Section \ref{summalize} provides a comprehensive discussion based on the DMD and the BMD results. The main results and findings are summarized in section \ref{conclude}, and the future directions are described.

\section{Numerical model and modal analysis}\label{numerical}
This section presents the numerical simulation methods for Mode A and the methodology of modal analysis applied to the numerical results.
\subsection{Numerical simulation method and computational grids}
The flow around a circular cylinder at $Re=200$ was obtained from a numerical simulation of the incompressible Navier--Stokes equations. The governing equations are presented below.
\begin{equation}
\nabla\cdot\boldsymbol{u}=0,
   \label{eqcont}	
\end{equation}
\begin{equation}
\frac{\partial{\boldsymbol{u}}}{\partial{t}}=-\nabla\cdot\boldsymbol{u}\boldsymbol{u}-\frac{1}{\rho}\nabla{p}+\frac{1}{Re}{\nabla^2}\boldsymbol{u}.
   \label{eqnavi}
\end{equation}
where $\boldsymbol{u}$ represents the velocity vector (bold symbols represent vectors), ${p}$ is the pressure, and $\rho$ is the fluid density. Here, the $Re$ is defined as
\begin{equation}
Re=\frac{U_{\infty}D}{\nu},
   \label{eqRe}
\end{equation}
where $U_{\infty}$ denotes the free-stream velocity, $\nu$ is the kinematic viscosity, and $D$ is the cylinder diameter. 

The governing equations are discretized based on the fractional step method proposed by \citet{RKincomp}. This method uses the third-order three-stage Runge--Kutta scheme for the advection term and the second-order implicit Crank--Nicholson scheme for the viscous term in the time advancement. The time step size was set such that the maximum Courant–Friedrichs–Lewy (CFL) number \citep{CFL} across all cells was less than or equal to $0.5$ in this paper. The validation for the time step size is provided in appendix \ref{appen1}. Spatial differences were evaluated using the second-order central difference \citep{poisson} and the QUICK method \citep{QUICK}. The pressure Poisson equation was solved using the bi-conjugate gradient stabilized method \citep{BiCGstab}. The details of these numerical procedures are described in the previous studies \citep{mypaper2, BBAA}. 

\begin{figure}
\begin{tabular}{cc}
\multicolumn{1}{l}{(\textit{a})}  &  \multicolumn{1}{l}{(\textit{b})}\\
  \begin{minipage}[b]{0.5\linewidth}
      %\subcaption{}
          \centering\includegraphics[width=6.5cm, keepaspectratio]{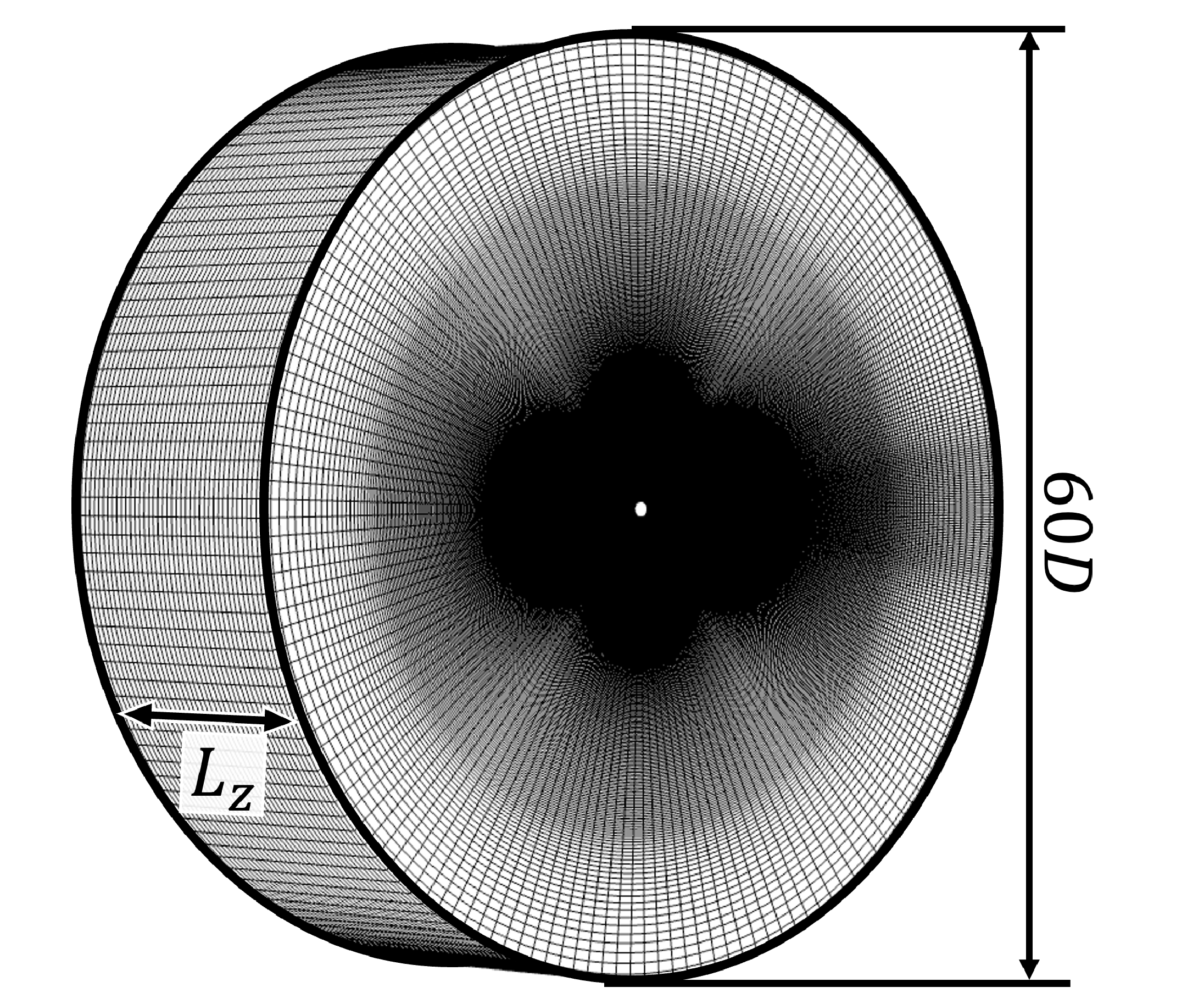}
    %\subcaption{Result at t = 0.25}
  \end{minipage}
  &
  \begin{minipage}[b]{0.5\linewidth}
      %\subcaption{}
          \centering\includegraphics[width=6.5cm, keepaspectratio]{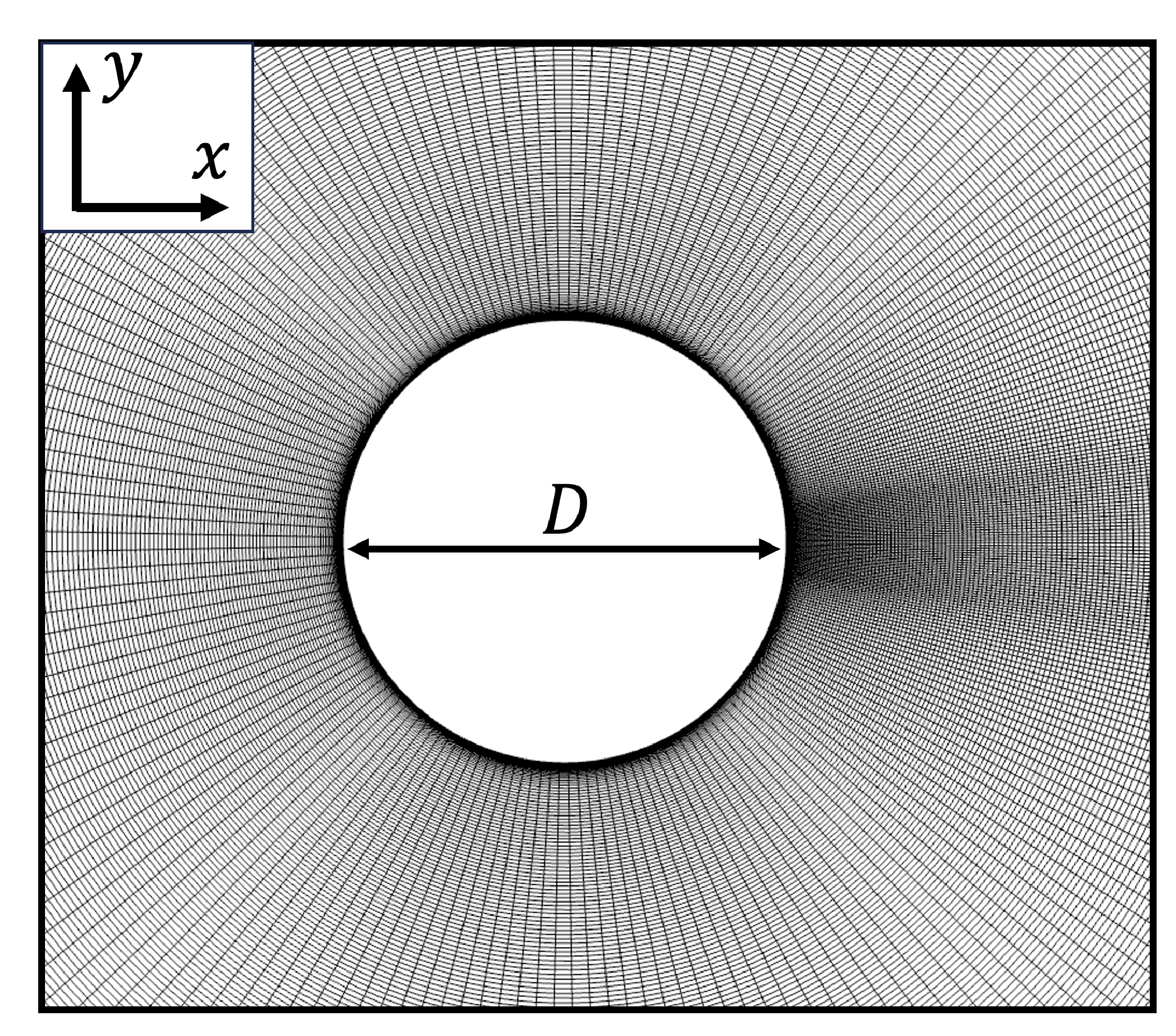}
    %\subcaption{Result at t = 0.5}
  \end{minipage}%\\
  \end{tabular}
    \captionsetup{justification=raggedright,singlelinecheck=false}
  \caption{Computational grid around a circular cylinder: (\textit{a}) overall grids (\textit{b}) close-up view of near the cylinder.}
 \label{fig:figure1}
\end{figure}

The three-dimensional computational mesh is shown in figure \ref{fig:figure1}. The far-field boundary of the computational domain extended up to $60$ times the diameter of the circular cylinder $D$. The number of cells was $240$ in the wall-normal direction and $440$ in the wall-parallel direction. The height of the first layer next to the cylinder was $1.0 \times 10^{-3}$. A periodic boundary condition was imposed on the spanwise boundary. The spanwise domain size $L_z$ varied by adjusting the number of grids, whereas the spanwise grid width $dz$ was fixed at $0.1D$. These grid parameters and boundary conditions were based on the DNS of \citet{modeA,cylinder_DNS,cylinder4}. 

\subsection{DMD algorithm}
The DMD is the means of decomposing an original flow field into a series of spatial modes. Each mode has a characteristic frequency and growth rate. Although several derivative algorithms exist for the DMD \citep{DMD3}, we introduce the DMD algorithm for the vast amount of time series data of the flow field obtained from the numerical simulation \citep{Hemati_2014,IPOD_Ohmichi}. 

Let $\boldsymbol{u}(\boldsymbol{x},t_j) (j=1,2, \cdots M)$ be the time series data of the velocity field, which is the vector of velocities in three directions $x$, $y$, $z$ of all grid points in a row, with $N$ elements. In the DMD, two data matrices $X$ and $X'$ composed by $\boldsymbol{u}(\boldsymbol{x}, t_j)$ are considered:
 \begin{equation}
  \begin{split}
X=[\boldsymbol{u}(\boldsymbol{x},t_1), \boldsymbol{u}(\boldsymbol{x},t_2), \cdots, \boldsymbol{u}(\boldsymbol{x},t_{M-1})] \in \mathbb{R}^{N \times M-1},\\
X'=[\boldsymbol{u}(\boldsymbol{x},t_2), \boldsymbol{u}(\boldsymbol{x},t_3), \cdots, \boldsymbol{u}(\boldsymbol{x},t_{M})] \,\,\,\,\,\,   \in \mathbb{R}^{N \times M-1}.
  \label{DMDmat}
  \end{split}
\end{equation}
These two matrices are related by matrix $A\in \mathbb{R}^{N \times N}$ as follows
\begin{equation}
  \begin{split}
X'=AX.
  \label{DMDmat}
  \end{split}
\end{equation}
The DMD aims to compute the eigenvalues and eigenvectors of matrix $A$. However, when the flow field is given by the numerical simulation, the size of the matrix $A$ is quite large, which makes it impractical to solve the eigenvalue problem directly. Therefore, the singular value decomposition (SVD) or proper orthogonal decomposition (POD)\cite{POD0} is applied to $X$ to reduce the dimension of the eigenvalue problem for matrix $A$. In the SVD, $X$ is represented by:
 \begin{equation}
X \approx U_rS_rV_r^{T},
  \label{SVD}
\end{equation}
where $U_r \in \mathbb{R}^{N \times r}$ and $V_r \in \mathbb{R}^{r \times M-1}$ represent the left and right singular vectors, respectively, $S_r \in \mathbb{R}^{r \times r}$ is the diagonal matrix with non-negative diagonal elements (the singular values of $X$), and $\cdot^T$ is the transpose of a matrix. In the POD context, $U_r$ is a matrix of the POD modes. The POD modes here are computed without subtracting the time-averaged field from the data set. The difference due to the presence or absence of time averages was referred to in the \citet{mfPOD}. The number of SVD modes $r$ was selected such that the cumulative contributions of the fluctuating components exceeded $99.5$ \%. Details of the rank determination method are provided in \citet{mfPOD}. 
To handle large data sets, this study uses incremental POD \citep{IDMD,Ross_2008,IPOD_Ohmichi} for computing the matrix $U_r$ consisting of $r$ POD modes.  Furthermore, the POD is parallelized by the idea of the \citet{asada2025exact}. From the matrices $X$ and $U_r$, the low-rank matrix $\tilde{X}$ is computed from
 \begin{equation}
\tilde{X}={U_r}^{T}X.
  \label{Aapproxomation}
\end{equation}

The low-rank approximation $\tilde{A}$ of the matrix $A$ can be computed as
 \begin{equation}
\tilde{A}={U_r}^TX'\tilde{X}^T(\tilde{X}\tilde{X}^T)^{\dagger},
  \label{Aapproximation}
\end{equation}
where the superscript $\dagger$ denotes the Moore–Penrose pseudoinverse. This approximation is based on the derivation of \citet{Hemati_2014}, and the right-hand side is mathematically coincident with ${U_r}^TX'V_rS_r^{-1}$. From the $k$th eigenvector $\boldsymbol{\tilde{\varphi}}^{\text{DMD}}_k$ of the low-rank matrix $\tilde{A}$, the $k$th DMD mode $\boldsymbol{\varphi}^{\text{DMD}}_k$ can be computed as:
 \begin{equation}
\boldsymbol{\varphi}^{\text{DMD}}_k = U_r\boldsymbol{\tilde{\varphi}}^{\text{DMD}}_k.
  \label{DMDmode}
\end{equation}
The frequency $f^{\text{DMD}}_k$ and the growth rate $\sigma^{\text{DMD}}_k$ of the $k$th DMD mode $\boldsymbol{\varphi}^{\text{DMD}}_k$ are
 \begin{equation}
f^{\text{DMD}}_k = \frac{\Imag\{\log({\tilde{\lambda}^{\text{DMD}}_k})\}}{\Delta t},
  \label{DMDfreq}
\end{equation}
 \begin{equation}
\sigma^{\text{DMD}}_k = \frac{\Real\{\log({\tilde{\lambda}^{\text{DMD}}_k})\}}{\Delta t},
  \label{DMDfreq}
\end{equation}
where $\Real(\cdot)$ and  $\Imag(\cdot)$ represent the real and imaginary parts of the complex values, respectively, and $\tilde{\lambda}^{\text{DMD}}$ is the eigenvalue corresponding to the $k$th DMD mode $\boldsymbol{\tilde{\varphi}}^{\text{DMD}}_k$. This paper adopts the Hankel matrix \citep{brunton2017chaos,asada2025exact} for $X$ and $X'$ to increase the accuracy of the computation results.

\subsection{BMD algorithm}
The BMD \citep{schmidt_2020} detects spatial structures that relate to triadic interaction in the statistically stationary flow field. The frequencies {$f_p,f_q,f_r$} of spatial structures in the triadic interaction are sum to zero:
 \begin{equation}
  \begin{split}
f_p \pm f_q \pm f_r=0.
  \label{sumtozero}
  \end{split}
\end{equation}
In the BMD context, the interaction relationship in terms of the bispectrum and its spatial structure are identified based on the quadratic term of Navier-Stokes equation.

%ここからBMDの説明
Let $\boldsymbol{u}(\boldsymbol{x},t_m) (m=1,2, \cdots M)$ be the time series data of the velocity field. Welch's method is adopted to estimate an asymptotically consistent power spectrum and bispectrum. The time series data is divided into a number of $N_{\text{blk}}$ segments. Each segment is composed of a number of $N_\text{FFT}$ snapshots, and the segments overlap by $N_{\text{ovlp}}$.  
 
The discretized Fourier transform of $l$th segment is represented by:
\begin{equation}
  \begin{split}
  \hat{\boldsymbol{u}}^{l}(\boldsymbol{x},f_p) = \sum^{N_\text{FFT}-1}_{m=0}\boldsymbol{u}(\boldsymbol{x},t_m)e^{-\frac{2\pi ipm}{N_\text{FFT}}}\,\,\,\,\,\,\,(p=0,1,\cdots N_\text{FFT}-1).
  \label{FFT}
  \end{split}
\end{equation}
The sampling frequency of $\hat{\boldsymbol{u}}$ is determined by $\frac{1}{\Delta t}$. Here, the bispectrum of frequency $f_p$ and $f_q$ is defined by
\begin{equation}
  \begin{split}
  \mathcal{B}(f_p,f_q) = E[\langle \hat{\boldsymbol{u}}^*(\boldsymbol{x},f_p) \circ \hat{\boldsymbol{u}}^*(\boldsymbol{x},f_q) , \hat{\boldsymbol{u}}(\boldsymbol{x},f_{p+q})\rangle]
  \label{FFT}
  \end{split}
\end{equation}
where $\circ$ denotes element-wise product. 
Note that subscript $r$ is equal to subscript $p+q$ when discretized frequencies are related to the sum to zero of equation \ref{sumtozero}.
From the point of the Navier-Stokes equation, $\hat{\boldsymbol{u}}(\boldsymbol{x},f_{p+q})$ is derived from the quadrastic term of $\hat{\boldsymbol{u}}^*(\boldsymbol{x},f_p)$ and $\hat{\boldsymbol{u}}^*(\boldsymbol{x},f_q)$ \citep{schmidt_2020,freeman_2024}. Therefore, bispectrum means the interaction relationship between $\hat{\boldsymbol{u}}^*(\boldsymbol{x},f_p) \circ \hat{\boldsymbol{u}}^*(\boldsymbol{x},f_q)$ (cause) and $\hat{\boldsymbol{u}}(\boldsymbol{x},f_{p+q})$ (effect). 

Here, bispectrum is computed from $N_\text{blk}$ segment spectrum. That is, the cause and effect of each block are coupled by the coefficient $a_k (f_p,f_q)\,\,\,(k=1, 2, \cdots N_{\text{blk}})$ as follows:
\begin{equation}
  \begin{split}
  \boldsymbol{\phi}_{p \circ q}({\boldsymbol{x},f_p,f_q})&=\sum_{k=1}^{N_{\text{blk}}}a_k(f_p,f_q)\{\hat{\boldsymbol{u}}^{k}(\boldsymbol{x},f_p) \circ \hat{\boldsymbol{u}}^{k}(\boldsymbol{x},f_q)\},\\
  \boldsymbol{\phi}_{p + q}({\boldsymbol{x},f_{p+q}})&=\sum_{k=1}^{N_{\text{blk}}}a_k(f_p,f_q)\{\hat{\boldsymbol{u}}(\boldsymbol{x},f_{p+q})\},
  \label{couple}
  \end{split}
\end{equation}
where $ \boldsymbol{\phi}_{p \circ q}({\boldsymbol{x},f_p,f_q})$ and $\boldsymbol{\phi}_{p + q}({\boldsymbol{x},f_{p+q}})$ represent the cause and effect in terms of bispectrum summarized on all segments.
The core of the BMD is detecting the coupling coefficient $a_k$ to maximize the bispectrum. Thus, $a_k$ is formulated by following the maximization problem:
\begin{equation}
  \begin{split}
    \boldsymbol{a} = \underset{\|\mathbf{a}\|=1} {\operatorname{argmax}}
    \left| E \left[\langle \boldsymbol{\phi}_{p \circ q} , \boldsymbol{\phi}_{p+q}\rangle \right] \right|
  \end{split}
  \label{max}
\end{equation}
This maximization problem can be solved by computing the eigenvector corresponding to the largest eigenvalue of the following matrix $B_{p,q}$:
\begin{equation}
  \begin{split}
  B_{p,q}=U_{p \circ q}^{H}&WU_{p + q} \in \mathbb{R}^{N_\text{blk} \times N_\text{blk}},\\
    U_{p \circ q}&=[\hat{\boldsymbol{u}}^1_{p \circ q}, \hat{\boldsymbol{u}}^2_{p \circ q}, \cdots   \hat{\boldsymbol{u}}^{N_\text{blk}}_{p \circ q}] \in \mathbb{R}^{N \times N_\text{blk}},\\
    U_{p + q}&=[\hat{\boldsymbol{u}}^1_{p + q}, \hat{\boldsymbol{u}}^2_{p + q}, \cdots  \hat{\boldsymbol{u}}^{N_\text{blk}}_{p + q}] \in \mathbb{R}^{N \times N_\text{blk}},
  \end{split}
  \label{max}
\end{equation}
where $\hat{\boldsymbol{u}}^k_{p \circ q}$ and $\hat{\boldsymbol{u}}^k_{p+q}$ are presented below:
\begin{equation}
  \begin{split}
  \hat{\boldsymbol{u}}^k_{p \circ q} &= \hat{\boldsymbol{u}}^{k}(\boldsymbol{x},f_p) \circ \hat{\boldsymbol{u}}^{k}(\boldsymbol{x},f_q),\\
  \hat{\boldsymbol{u}}^k_{p+q}&=\hat{\boldsymbol{u}}(\boldsymbol{x},f_{p+q}).
  \label{couple}
  \end{split}
\end{equation}
and the maximum eigenvalue $\lambda^{\text{BMD}}_{p,q}$ is equal to the bispectrum. We compute the eigenvalue and its eigenvector of $B_{p,q}$ using the ZGEEV routine implemented in the lapack library. 

\begin{figure}
    \centering\includegraphics[keepaspectratio, scale=0.7]{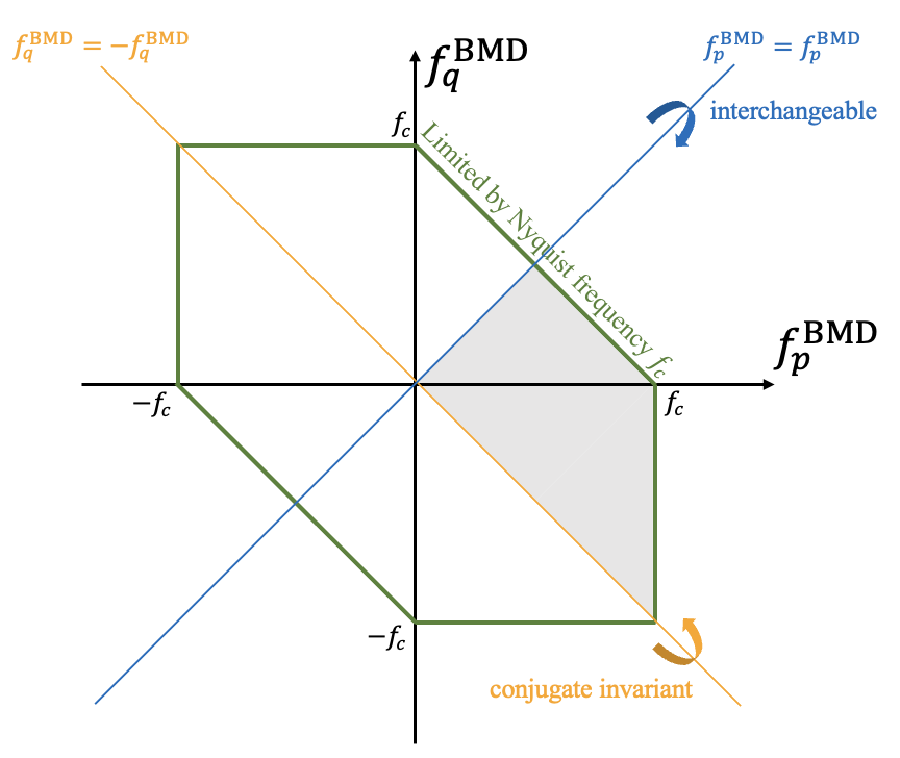}
    \captionsetup{justification=raggedright,singlelinecheck=false}
  \caption{Symmetricity of bispectrum. Due to the Nyquist frequency limitation, conjugate invariant, and symmetry of $f^{\text{BMD}}_p$ and $f^{\text{BMD}}_q$, the bispectrum is computed only in the gray region.}
 \label{fig:figure_BMDregion}
\end{figure}

The bispectrum can be computed over arbitrary two frequencies $f^{\text{BMD}}_p$, $f^{\text{BMD}}_q$. However, depending on the Nyquist frequency $f_c$, it can be computed only in the region satisfying $-f_c\leq f^{\text{BMD}}_p \leq f_c$, $-f_c\leq f^{\text{BMD}}_q \leq f_c$, and $-f_c\leq f^{\text{BMD}}_p + f^{\text{BMD}}_q \leq f_c$. In addition, $\lambda^{\text{BMD}}_{p,q}$ has the same value even when $f^{\text{BMD}}_p$ and $f^{\text{BMD}}_q$ are interchanged or the bispectrum is conjugated. Therefore, $\lambda^{\text{BMD}}_{p,q}$ can be computed in the gray region in figure \ref{fig:figure_BMDregion} (principal region).

In the context of the BMD, the cause distribution $\boldsymbol{\phi}_{p \circ q}$ and its effect $\boldsymbol{\phi}_{p + q}$ is referred to as cross-frequency field and bispectral mode, respectively. The interaction relationship between $\boldsymbol{\phi}_{p \circ q}$ and $\boldsymbol{\phi}_{p + q}$ is characterized by the spatial mode defined below:
\begin{equation}
  \boldsymbol{\tau}_{p \circ q}(\boldsymbol{x}) = 
  \text{Abs}\{\boldsymbol{\phi}_{p \circ q}(\boldsymbol{x}) \circ \boldsymbol{\phi}_{p + q}(\boldsymbol{x})\},
  \label{interactionmap}
\end{equation}
where $\text{Abs}(\cdot)$ represents absolute value of complex number. The $\boldsymbol{\tau}_{p \circ q}$ represents the spatial distribution of interaction strength since the spatial integration of $\boldsymbol{\tau}_{p \circ q}$ is equal to $\text{Abs}(\lambda^{\text{BMD}}_{p,q})$. Therefore, $\boldsymbol{\tau}_{p \circ q}$ is referred to as an interaction map.

\section{Simulation results and modal analysis}\label{results}
\subsection{Numerical simulation results}
\begin{figure}
\begin{tabular}{cc}
\multicolumn{1}{l}{(\textit{a})}  &  \multicolumn{1}{l}{(\textit{b})}\\
  \begin{minipage}[b]{0.5\linewidth}
      %\subcaption{}
          \centering\includegraphics[width=6.5cm, keepaspectratio]{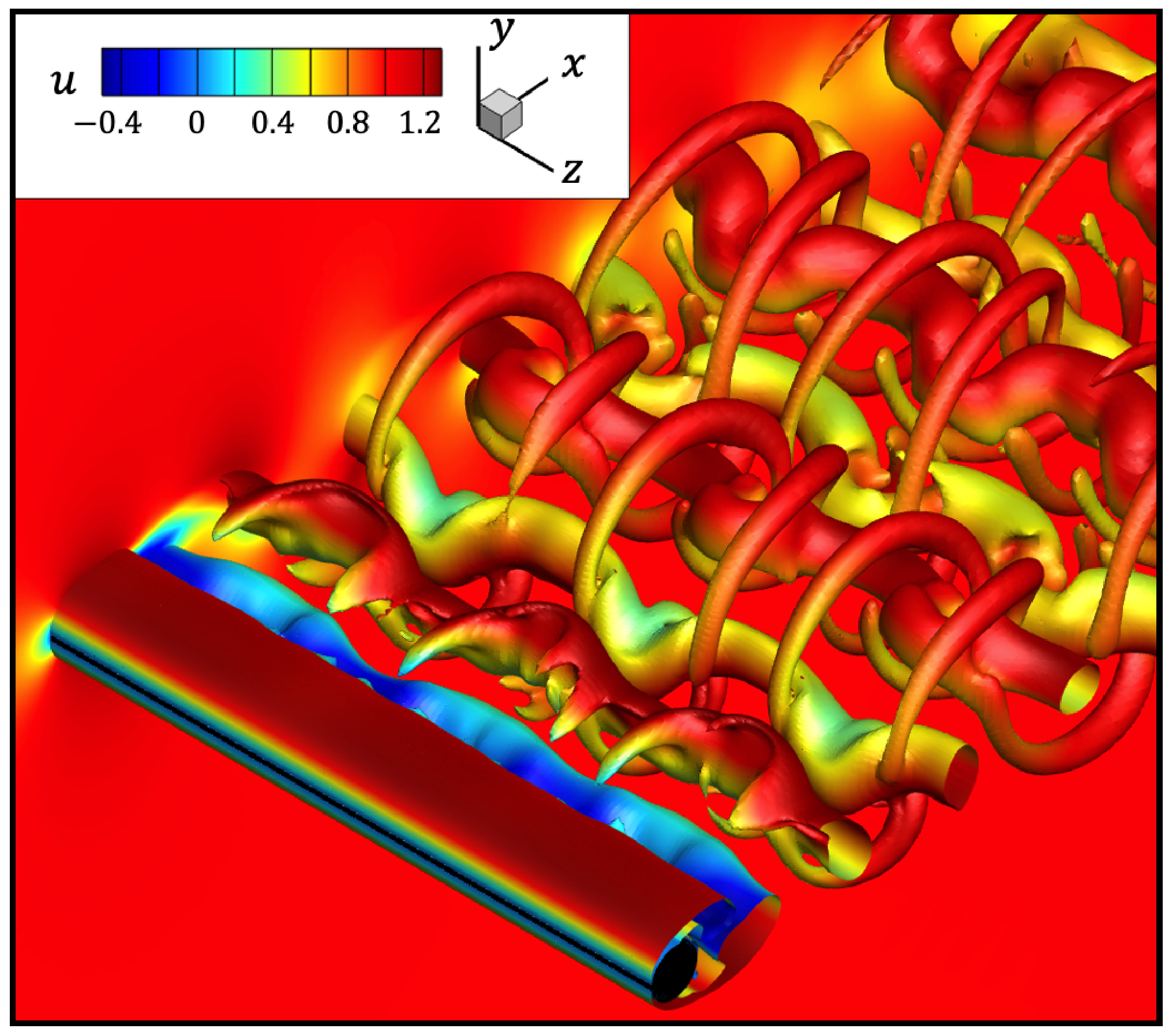}
    %\subcaption{Result at t = 0.25}
  \end{minipage}
  &
  \begin{minipage}[b]{0.5\linewidth}
      %\subcaption{}
          \centering\includegraphics[width=6.5cm, keepaspectratio]{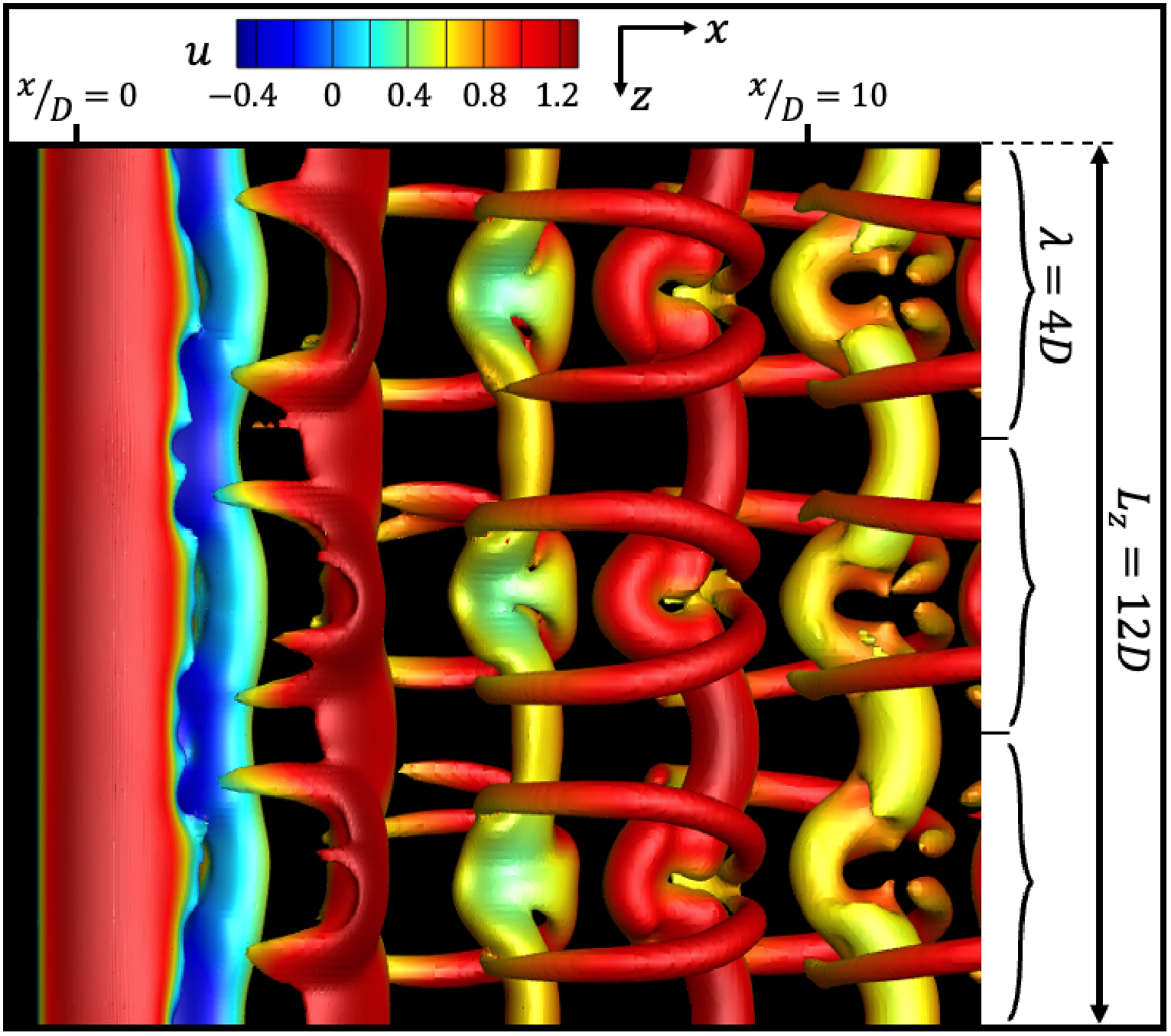}
    %\subcaption{Result at t = 0.5}
  \end{minipage}%\\
  \end{tabular}
    \captionsetup{justification=raggedright,singlelinecheck=false}
  \caption{Isosurface of Q-value at $0.1$ colored by $x$-drection velocity obtained in the computational domain of $L_z=12D$: (\textit{a}) overall flow fields (\textit{b}) $y$-normal views. The flow field has a three-dimensional vortex structure. Wavelength of spanwise direction is $4D$.}
 \label{fig:figure2}
\end{figure}

Numerical simulations were performed for $L_z=12D$. 
%According to Jiang et al.\cite{Jiang_2017}, different flow fields are formed depending on the value of $L_z$. 
According to \citet{Jiang_2017}, when $L_z\geq12D$, the average drag coefficient, root-mean-square of lift coefficient, and $St$ of Karman vortex shedding were almost the same. Therefore, the numerical simulation result at $L_z=12D$ was less affected by the spanwise domain size. Figure \ref{fig:figure2} shows the isosurface of Q-value colored by $x$-direction velocity at $L_z=12D$ obtained from the numerical simulation. The vortex structure behind the cylinder is periodic in the spanwise direction. The wavelength $\lambda$ at $L_z=12D$ was approximately $4D$, because three periodic structures were observed in the spanwise direction. This wavelength was consistent with those reported in \citet{Jiang_2017, modeA}.

According to \citet{Jiang_2017}, the wavelength $\lambda$ in the spanwise direction and the temporal behavior of force coefficients changed with $L_z$. To investigate the temporal behavior shift depending on the $L_z$, numerical simulations were performed for various $L_z$. Based on \citet{Jiang_2017,Rolandi_2023,cylinder4,mypaper_steady}, the time variation of the drag coefficient acting on the cylinder is a useful choice for characterizing the temporal behavior of the flow around the cylinder. The drag coefficient $C_D$ was computed from the numerical simulation results. This coefficient is computed as follows:
\begin{align}
C_D = \frac{1}{\frac{1}{2}\rho{U}_{\infty}^2DL_z}\int\int_\Omega(p\text{cos}\theta-\frac{1}{Re}{\omega}_z\text{sin}\theta)\,d\theta dz,
%C_L = \frac{1}{\frac{1}{2}\rho{U}_{\infty}^2DL_z}\int\int_\Omega(-p\text{sin}\theta-\frac{1}{\text{Re}}{\omega}_z\text{cos}\theta)\,d\theta dz,
\end{align}
where $\Omega$ denotes the cylinder surface, $\theta$ is the angle from the stagnation point, and the vorticity component along the $z$-axis, $\omega_z$, is calculated as follows: 
\begin{equation}
\omega_z = \frac{\partial v}{\partial x} - \frac{\partial u}{\partial y}.
   \label{eqvolz}
\end{equation}

\begin{figure}
\begin{tabular}{c}
\multicolumn{1}{l}{(\textit{a})}\\
  \begin{minipage}[b]{1\linewidth}
      %\subcaption{}
          \centering\includegraphics[width=13cm, keepaspectratio]{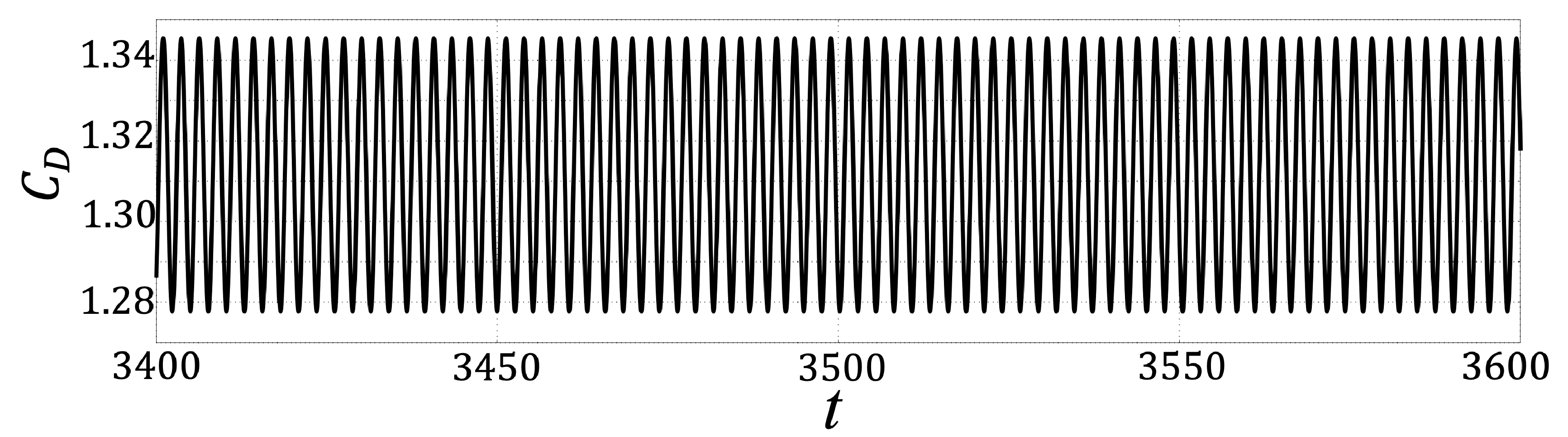}
    %\subcaption{Result at t = 0.25}
  \end{minipage}\\
\multicolumn{1}{l}{(\textit{b})}\\
  \begin{minipage}[b]{1\linewidth}
      %\subcaption{}
         \centering\includegraphics[width=13cm, keepaspectratio]{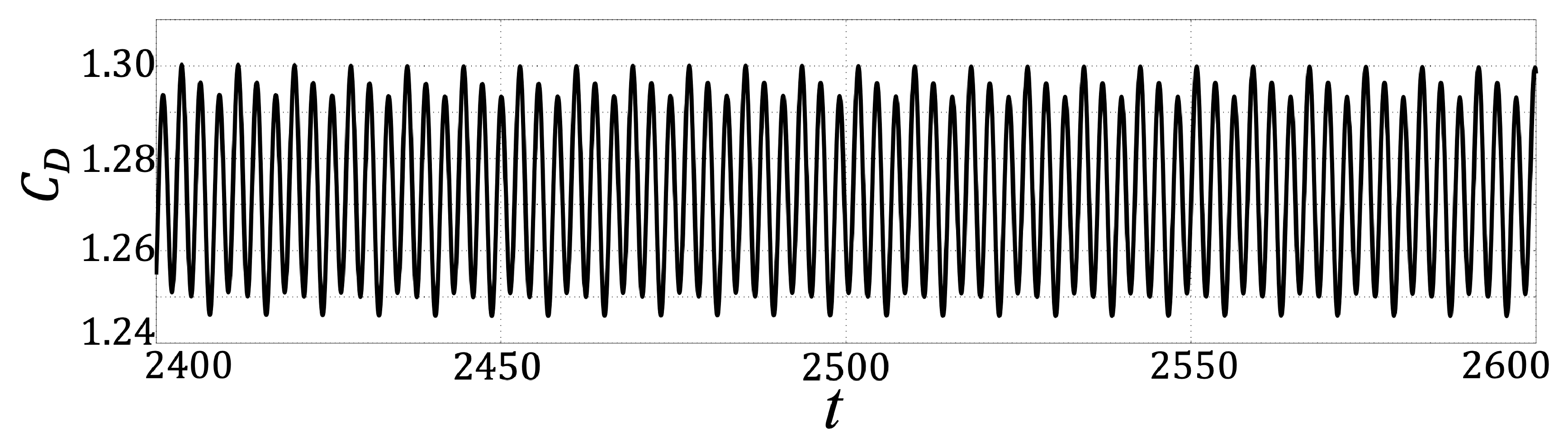}
    %\subcaption{Result at t = 0.5}
  \end{minipage}\vspace{-5pt}\\
\multicolumn{1}{l}{(\textit{c})}\\
  \begin{minipage}[b]{1\linewidth}
      %\subcaption{}
          \centering\includegraphics[width=13cm, keepaspectratio]{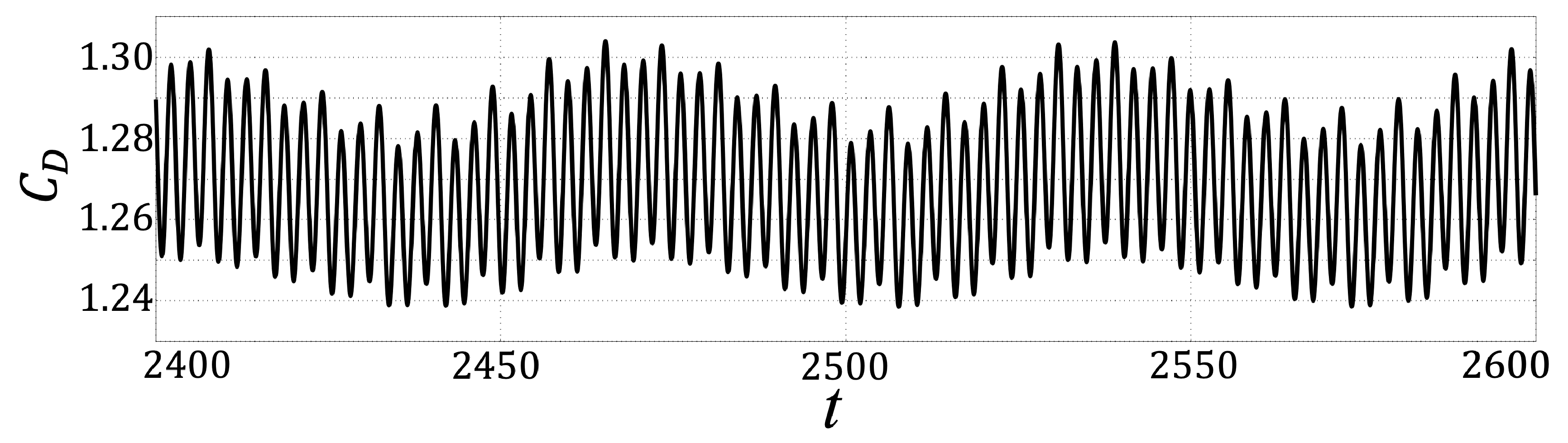}
    %\subcaption{Result at t = 0.5}
  \end{minipage}\\
\multicolumn{1}{l}{(\textit{d})}\\
    \begin{minipage}[b]{1\linewidth}
      %\subcaption{}
          \centering\includegraphics[width=13cm, keepaspectratio]{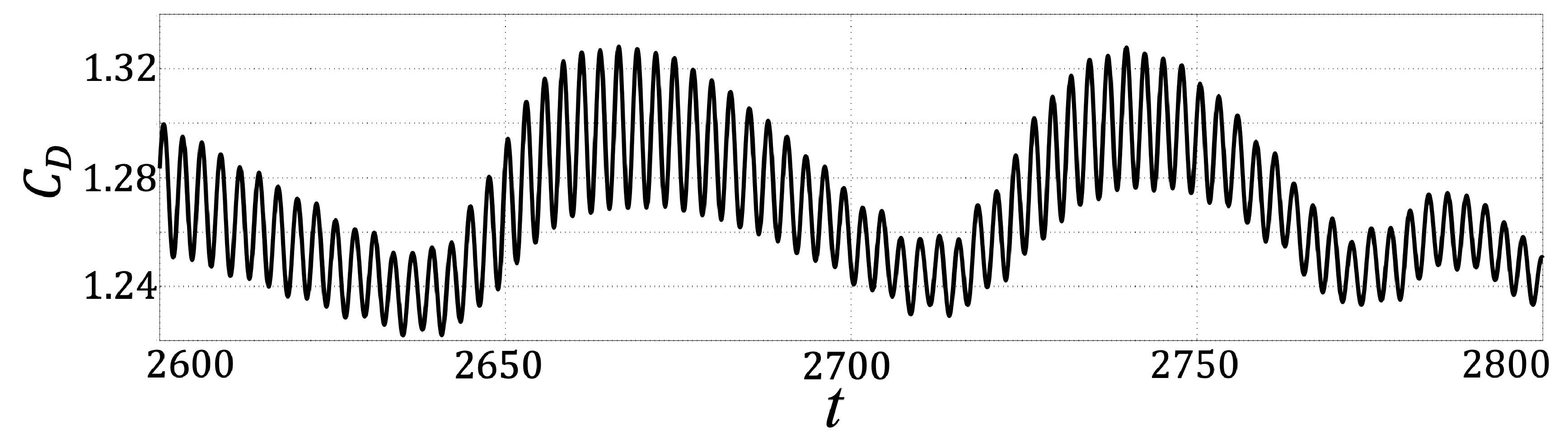}
    %\subcaption{Result at t = 0.5}
  \end{minipage}
  \end{tabular}
    \captionsetup{justification=raggedright,singlelinecheck=false}
  \caption{Time variation of drag coefficient with different spanwise domain size:   (\textit{a}) $L_z=3.4D$, (\textit{b}) $L_z=3.7D$, (\textit{c}) $L_z=3.9D$, and (\textit{d}) $L_z=12D$. As $L_z$ increases, low-frequency fluctuation becomes noticeable.}
 \label{fig:figure4}
\end{figure}

 Figure \ref{fig:figure4} shows the time variation in the drag coefficient with $L_z=3.4D, 3.7D, 3.9D,$ and $12D$, which exhibits the characteristic temporal behavior of the drag coefficient. At $L_z=3.4D$, the amplitudes of the drag coefficients were constant, which differs from the case where $L_z=12D$. This temporal property resembled the well-known two-dimensional periodic flow field around a circular cylinder \citep{BBAA}. When $L_z=3.7D$, there were slight and low-frequency fluctuations in the amplitude. However, the fluctuation was smaller than that for $L_z=12D$. When $L_z=3.9D$, the presence of a low-frequency fluctuation was inferred, as in the case of $L_z=12D$. Henderson denoted that the existence of multiple frequencies makes the low-frequency fluctuation \citep{Henderson}. The fluctuations in our results also indicate that multiple frequencies were present in the signal. 
 
The shift in low-frequency aspects owing to the spanwise domain size clearly appears in the time variation of the drag coefficient. When $L_z$ was small, the flow field was constrained by the spanwise domain size. In summary, the restriction of the wavelength in the spanwise direction by the spanwise domain size affects the temporal properties of the cylinder flow. However, a more detailed analysis is required to determine the frequency values in each $L_z$ case and the relationship between the different frequency components.
 
\subsection{DMD-based analysis}\label{DMD}
The spanwise boundary size affects the temporal behavior of the drag coefficient. Particularly, low-frequency beating occurs as the size of the boundary increases. To identify the spatial structures that oscillate at low frequencies, DMD is applied to the flow fields of various $L_z$. 

\begin{figure}
  \centering\includegraphics[keepaspectratio, scale=0.25]{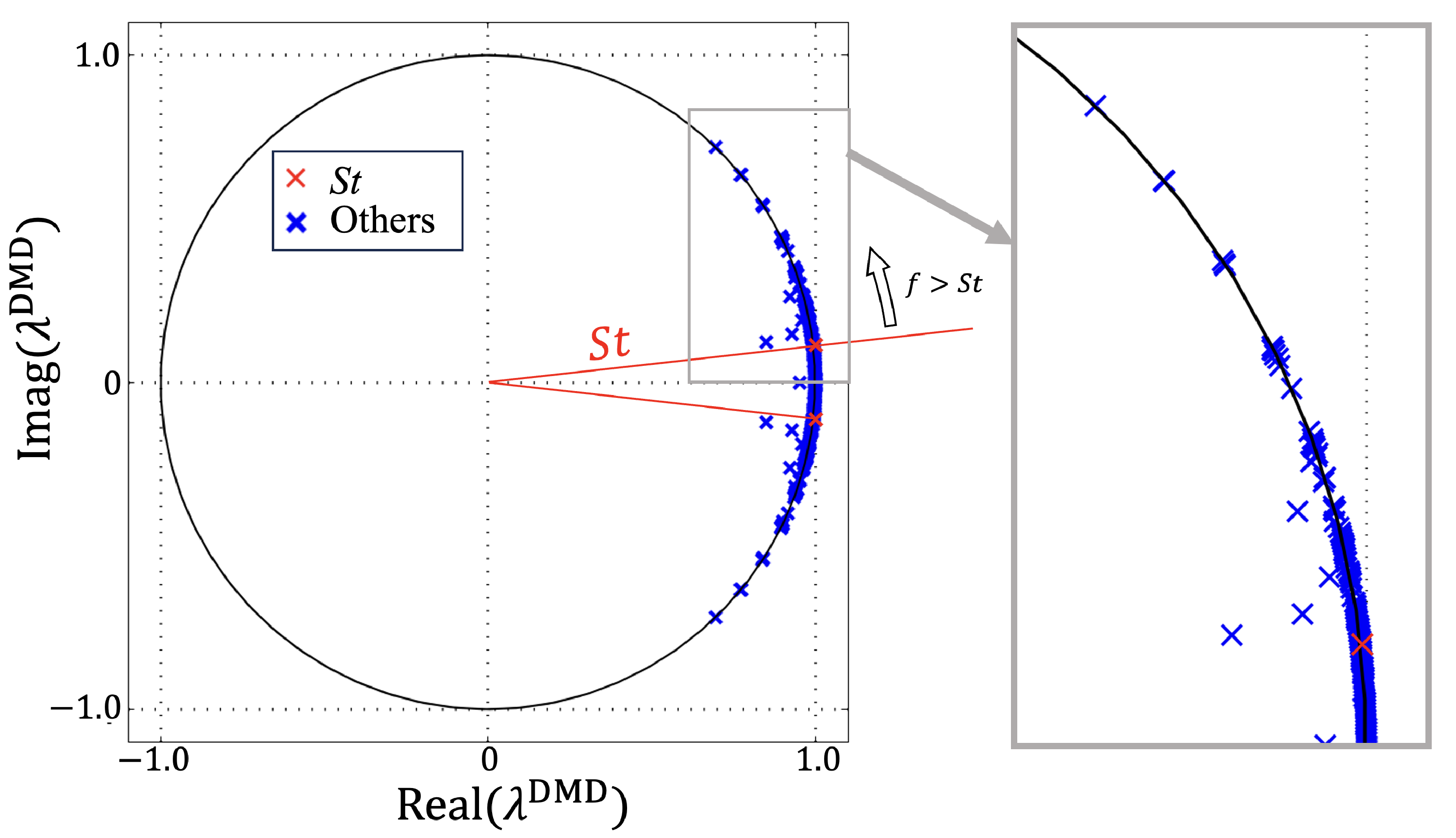}
    \captionsetup{justification=raggedright,singlelinecheck=false}
  \caption{Eigenvalues of DMD mode obtained from flow fields at $L_z=12D$. The SVD rank $r$ was $194$, which is determined based on the cumulative contribution ratio from SVD singular values for the fluctuating component.}
 \label{fig:figure3}
\end{figure}

DMD was performed for the time-series data of the flow fields in the $L_z=12D$ case.  In this case, the number of snapshots is $3000$, and the number of modes $r$ in the low-rank approximation of SVD was $194$, determined based on the cumulative contribution ratio exceeding $99.5 \%$. Figure \ref{fig:figure3} shows the eigenvalue distribution of the DMD mode. Here, the all frequency is  non-dimensionalized by $D$ and $U_{\infty}$, and frequency of Karman vortex shedding $St$ is defined as
\begin{equation}
St=\frac{fD}{U_{\infty}},
   \label{eqSt}
\end{equation}
where $f$ denotes the most dominant frequency of the cylinder wake. 
The eigenvalues on the unit circle have no growth rate and correspond to the stable modes. In this study, all the eigenvalues ideally existed in a unit circle since the fully developed quasi-steady flow dataset was used. However, some eigenvalues were located inside the unit circle. The modes inside the unit circle represent the damping mode. Such damping modes capture the frequency component in a portion of the statistical time for the datasets and are not important modes in the flow fields.

Focusing on the unit circle, the most dominant frequency $St$ selected by the greedy algorithm \citep{IDMD} is $0.1857$, which is consistent with \citet{Jiang_2017,cylinder4}. In the frequencies other than $St$, many frequencies exist that are lower than $St$. These low-frequency components correspond to the low-frequency fluctuation of the drag coefficient in figure \ref{fig:figure4}. However, due to the existence of a large number of low-frequency components in the $L_z=12D$ case, their selection and analysis are challenging.

The time variation of the drag coefficient shown in figure \ref{fig:figure4} is expected to suppress the appearance of low-frequency components and simplify the flow due to the constraint of the spanwise domain size. Thus, applying DMD to these $L_z$ cases provides a clear view. In the DMD mode computation process, the rank in the SVD approximation $r$ is truncated based on the cumulative contribution ratio exceeding $99.8 \%$. The truncated rank in the case of $L_z=3.4D, 3.7D,$ and $3.9D$ for the cases with figure \ref{fig:figure4} is shown in table \ref{table_SVD}. The increase in $r$ with increasing $L_z$. This means high-ranked modes (corresponding to small singular values) also have relatively high contribution rates, and many frequency components can be obtained from the DMD in the large $L_z$ case.

\begin{table}
 \centering
  \begin{tabular}{cccc}
   $L_z$ & $3.4D$ & $3.7D$ & $3.9D$\\[3pt]
   $r$ & $11$ & $31$ & $46$ \\
  \end{tabular}
    \captionsetup{justification=raggedright,singlelinecheck=false}
   \caption{The rank of SVD approximation in the DMD computation determined by the cumulative contribution ratio exceeding $99.8 \%$.}
 \label{table_SVD}
\end{table}

 \begin{figure}
\begin{tabular}{ccc}
\multicolumn{1}{l}{(\textit{a})}  &  \multicolumn{1}{l}{(\textit{b})} &  \multicolumn{1}{l}{(\textit{c})}\\
  \begin{minipage}[b]{0.32\linewidth}
      %\subcaption{}
          \centering\includegraphics[height=7cm,keepaspectratio]{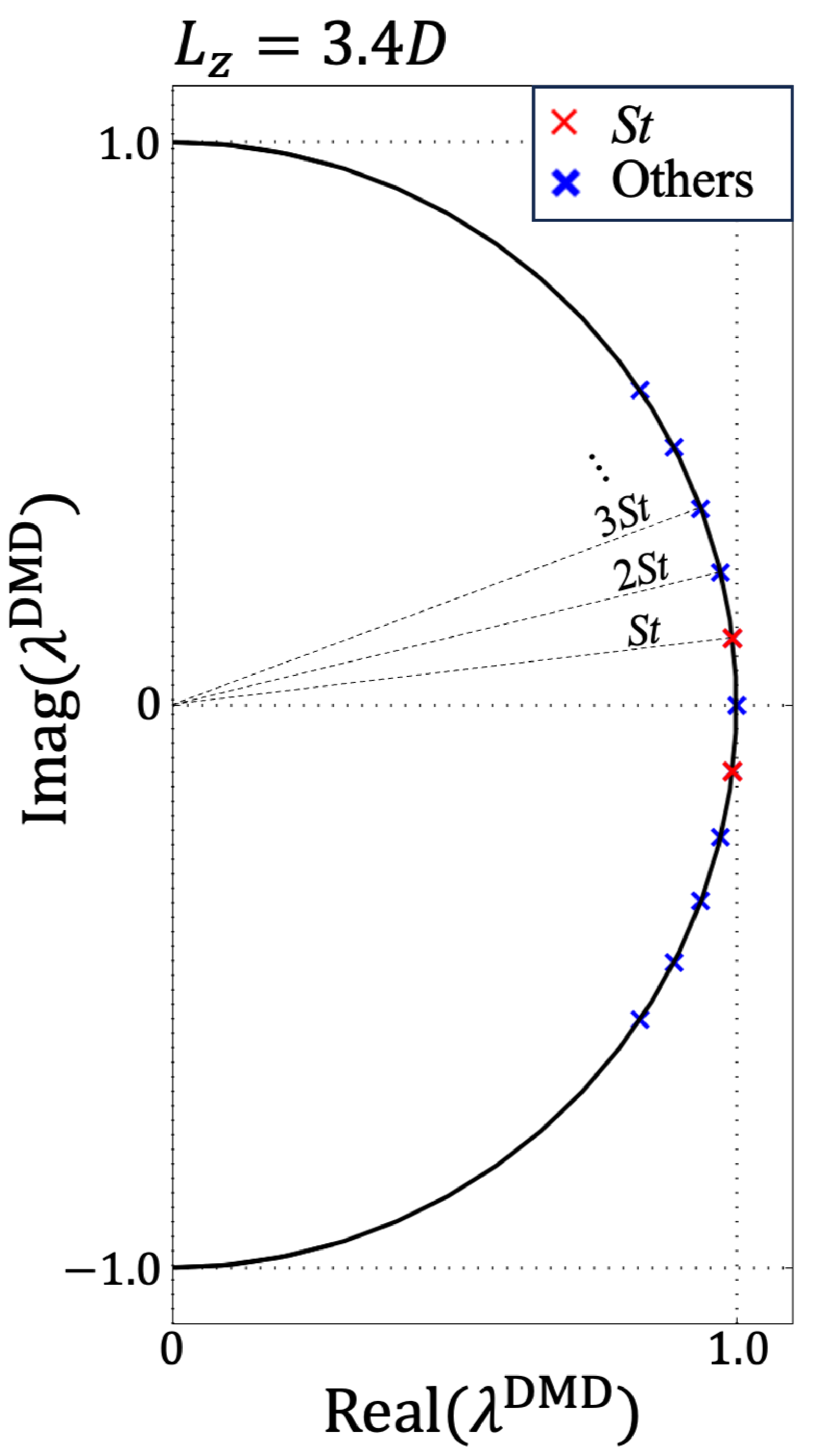}
    %\subcaption{Result at t = 0.25}
  \end{minipage}
  &
  \begin{minipage}[b]{0.32\linewidth}
      %\subcaption{}
          \centering\includegraphics[height=7cm,keepaspectratio]{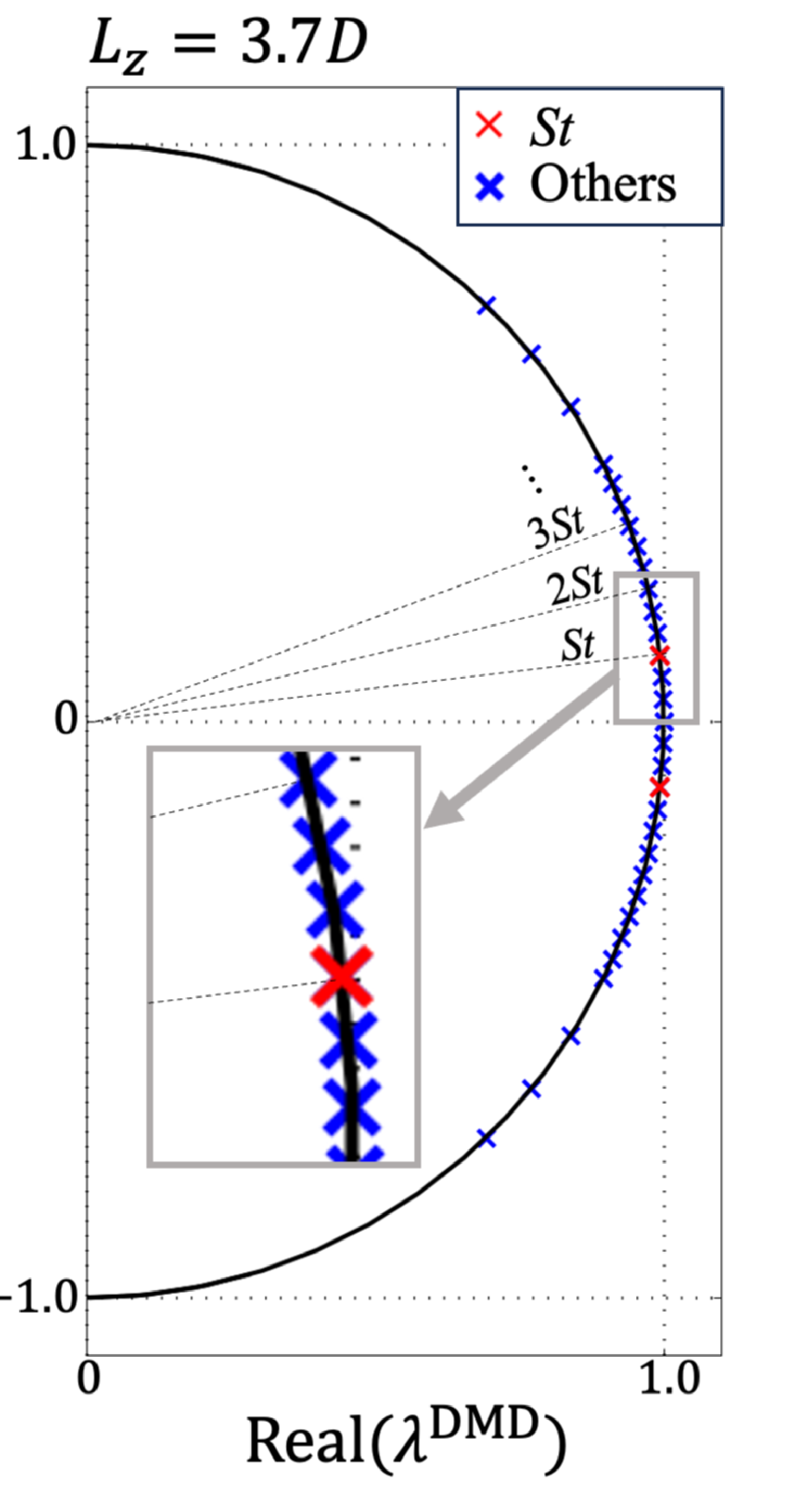}
    %\subcaption{Result at t = 0.5}
  \end{minipage}%\\
    &
  \begin{minipage}[b]{0.32\linewidth}
      %\subcaption{}
          \centering\includegraphics[height=7cm,keepaspectratio]{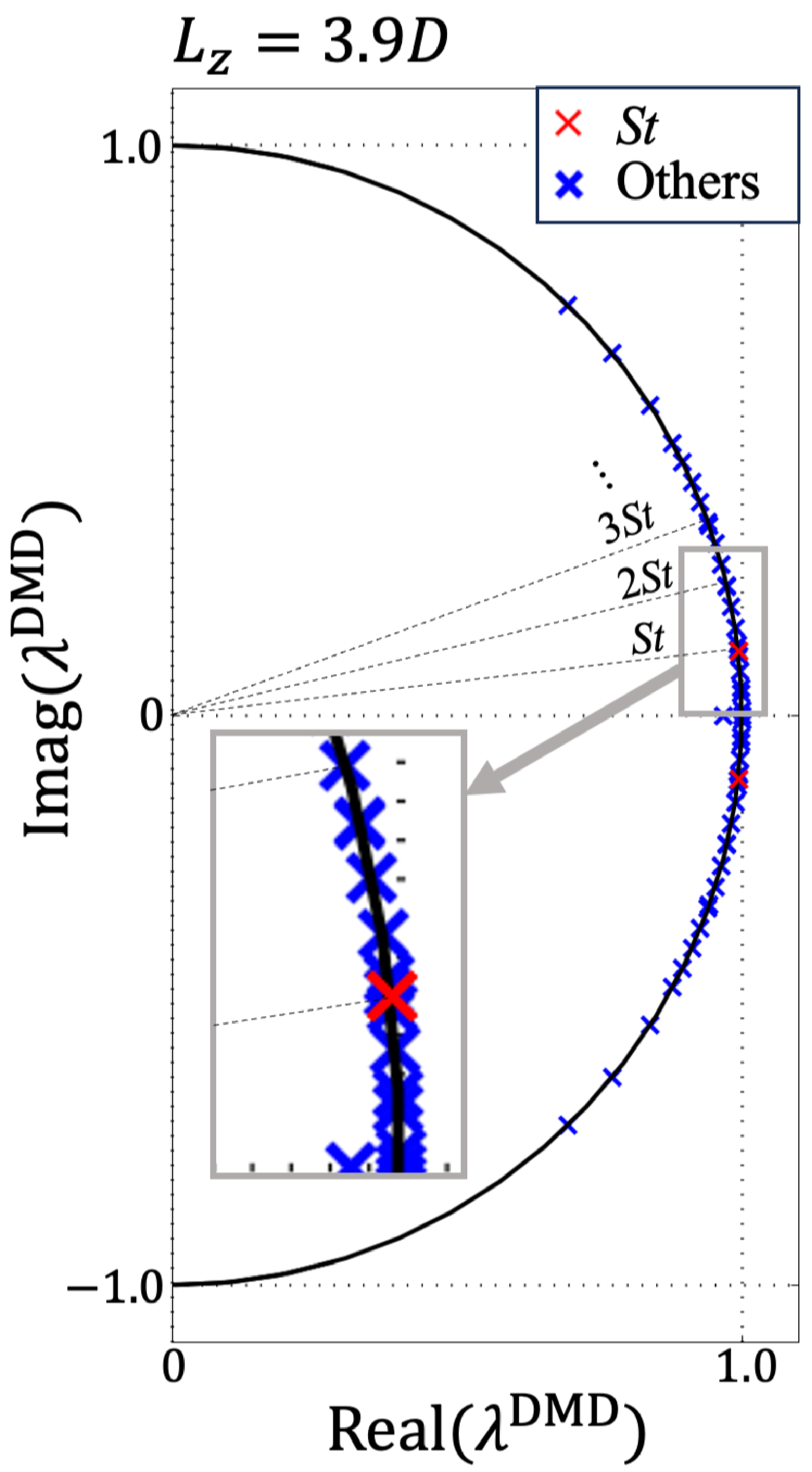}
    %\subcaption{Result at t = 0.5}
  \end{minipage}%\\
  \end{tabular}
    \captionsetup{justification=raggedright,singlelinecheck=false}
  \caption{Eigenvalues of DMD mode obtained from flow fields at (\textit{a}) $L_z=3.4D$, (\textit{b}) $L_z=3.7D$, and (\textit{c}) $L_z=3.9D$. The rank of SVD approximation is shown in table \ref{table_SVD}. In the case of $L_z=3.4D$, only the harmonics of $St$ are confirmed. For $L_z \geq 3.7D$ and $3.9D$, there exist the eigenvalues of a frequency other than the harmonics of $St$. }
 \label{fig:figure31}
\end{figure}

Figure \ref{fig:figure31} shows the eigenvalue distribution of DMD modes in the same $L_z$ cases for table \ref{table_SVD} and figure \ref{fig:figure4}. For $L_z=3.4D$ cases, no frequency lower than $St$ exists, and all frequencies are distributed at equal intervals. This means only harmonics of $St$ are identified. This is the same property as the flow around a two-dimensional cylinder, and the $L_z=3.4D$ case is completely periodic. In the $L_z=3.7D$ and $L_z=3.9D$, harmonics of $St$ are also identified. However, the frequency lower than $St$ becomes significant. That is, a low-frequency component exists in Mode A. At $L_z=3.7D$, all frequencies are distributed at equal intervals, which is common to $3.4D$. This implies the possibility that only harmonics of the lowest frequency exist in the $L_z=3.7D$ case. The lowest frequency, in this case, is about $0.06 \approx St/3$. Hereafter, we refer to this frequency component as ``third-subharmonics,'' while exactly one-third of $St$ is discussed in section \ref{BMD}.

Comparing the $3.7D$ and $3.9D$ cases, the number of frequencies smaller than $St$ differs. Furthermore, the $3.9D$ case has a clearly lower frequency than the $3.7D$ case. Hence, the appearance of the lowest frequency of $3.9D$ increases the number of frequency components lower than $St$. The shift with respect to $L_z$ characterized by the appearance of low-frequency components at the time variation of the drag coefficient in figure \ref{fig:figure4} was detected by the change in DMD eigenvalue distribution.

Based on the eigenvalue distribution of the DMD, we selected the DMD modes with $St$ in the $L_z=3.4D$ case and frequencies lower than $St$, the two lower frequency components $f^{\text{DMD}}_k = St/3$ and $2St/3$ in the $L_z=3.7D$ case, and the lowest frequency component $f^{\text{DMD}}_k \approx 0.01$ in the $L_z=3.9D$ case. Figure \ref{fig:figure31} shows the spatial distribution of the DMD modes for the four selected frequencies. The DMD mode of the $St$ represents the Karman vortex formed in the cylinder wake. We have confirmed this is common for all $L_z$ cases, but we have not shown it in the figure for simplicity.

 \begin{figure}
\begin{tabular}{cc}
\multicolumn{1}{l}{(\textit{a})}  &  \multicolumn{1}{l}{(\textit{b})}\\
  \begin{minipage}[b]{0.5\linewidth}
      %\subcaption{}
          \centering\includegraphics[width=6cm,keepaspectratio]{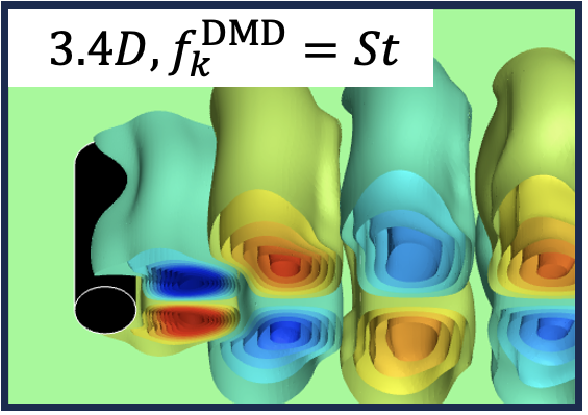}
    %\subcaption{Result at t = 0.25}
  \end{minipage}
  &
  \begin{minipage}[b]{0.5\linewidth}
      %\subcaption{}
          \centering\includegraphics[width=6cm,keepaspectratio]{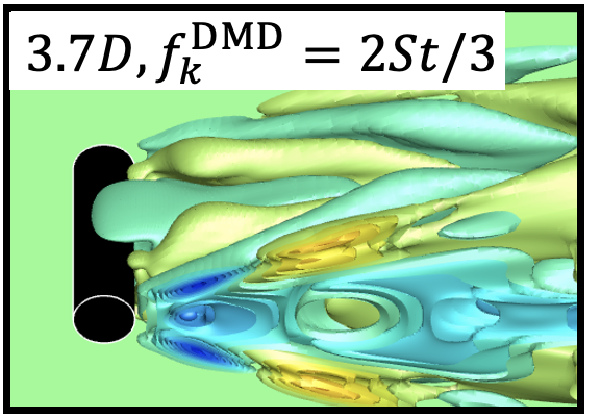}
    %\subcaption{Result at t = 0.5}
  \end{minipage}\\
  \multicolumn{1}{l}{(\textit{c})}  &  \multicolumn{1}{l}{(\textit{d})}\\
  \begin{minipage}[b]{0.5\linewidth}
      %\subcaption{}
          \centering\includegraphics[width=6cm,keepaspectratio]{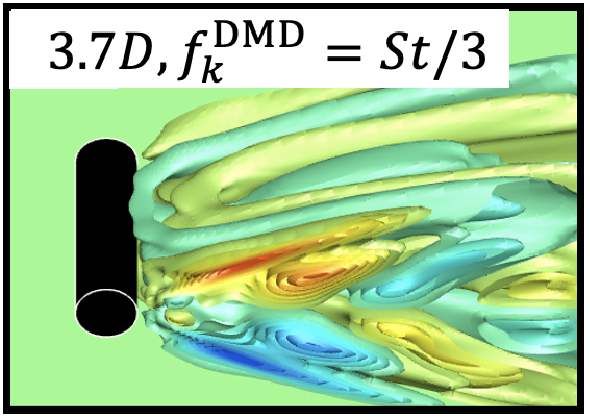}
    %\subcaption{Result at t = 0.25}
  \end{minipage}
  &
  \begin{minipage}[b]{0.5\linewidth}
      %\subcaption{}
          \centering\includegraphics[width=6cm,keepaspectratio]{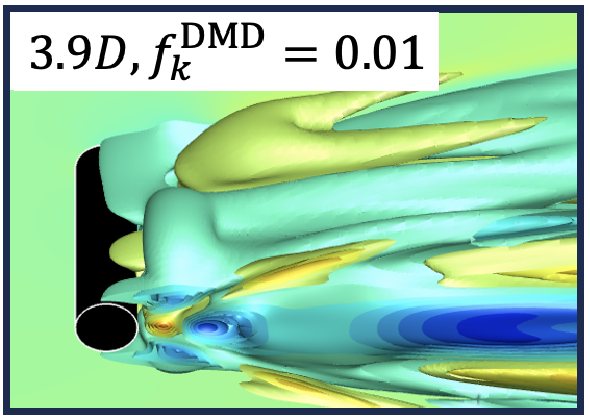}
    %\subcaption{Result at t = 0.5}
  \end{minipage}
   \end{tabular}
    \captionsetup{justification=raggedright,singlelinecheck=false}
  \caption{The real part of the $x$-direction component for the DMD mode. (\textit{a}) represents $f^{\text{DMD}}_k = St$ mode, (\textit{b}) is $f^{\text{DMD}}_k = 2St/3$ mode, (\textit{c}) is the $f^{\text{DMD}}_k = St/3$ mode, and (\textit{d}) is $f^{\text{DMD}}_k \approx 0.01$ mode. (\textit{a}) is the $L_z=3.4D$ case; (\textit{b}) and (\textit{c}) are the $L_z=3.7D$ case; and (\textit{d}) is the $L_z=3.9D$ case. (\textit{b}) and (\textit{c}) are selected because these DMD modes have a lower frequency than $St$ in the case of $3.7D$. (\textit{d}) is the lowest frequency DMD mode in $3.9D$ case.}
 \label{fig:figure32}
\end{figure}

The DMD modes of $f^{\text{DMD}}_k = 2St/3$ have the distribution of extending vertically toward the back of the cylinder. The Karman vortex of $St$ also showed a similar expansion in the backward direction. Like $f^{\text{DMD}}_k \approx = 2St/3$, $f^{\text{DMD}}_k = St/3$ has similar distribution behind the cylinder. However, the asymmetric structure along the $x$-axis in the cylinder wake differed from that in $f^{\text{DMD}}_k = 2St/3$. This asymmetry is the same as that of the Karman vortex, implying that $f^{\text{DMD}}_k = St/3$ is more similar to the oscillations of the Karman vortex than $f^{\text{DMD}}_k = 2St/3$. 

The DMD mode $f^{\text{DMD}}_k \approx 0.01$ has the different distribution from $f^{\text{DMD}}_k \approx St/3, 2St/3,$ and $St$. The symmetric structure of the cylinder wake resembles the well-known recirculation region that forms in the wake. At high $Re$ region, a phenomenon referred to as a ``bubble pumping'' was observed in the flow around various objects \citep{NAJJAR_1998,Yokota_2024,Yokota_2025,ohmichi2019numerical}. The DMD mode $f^{\text{DMD}}_k \approx 0.01$ in this study was similar to the structures identified in previous studies. Hence, the low-frequency fluctuation in Mode A is related to the bubble pumping.

\begin{figure}
\begin{tabular}{cccc}
\multicolumn{1}{l}{(\textit{a})}  &  \multicolumn{1}{l}{(\textit{b})}\\
  \begin{minipage}[b]{0.5\linewidth}
      %\subcaption{}
          \centering\includegraphics[width=6cm,keepaspectratio]{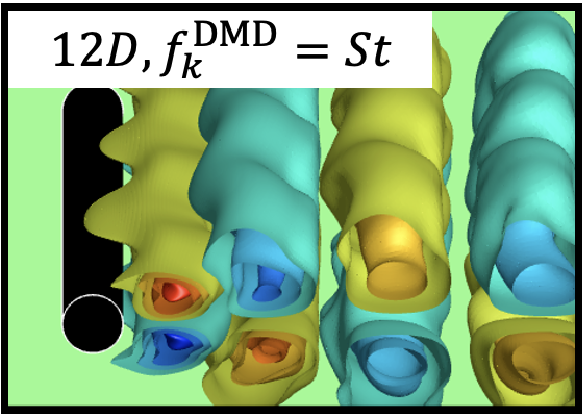}
    %\subcaption{Result at t = 0.25}
  \end{minipage}
  &
  \begin{minipage}[b]{0.5\linewidth}
      %\subcaption{}
          \centering\includegraphics[width=6cm,keepaspectratio]{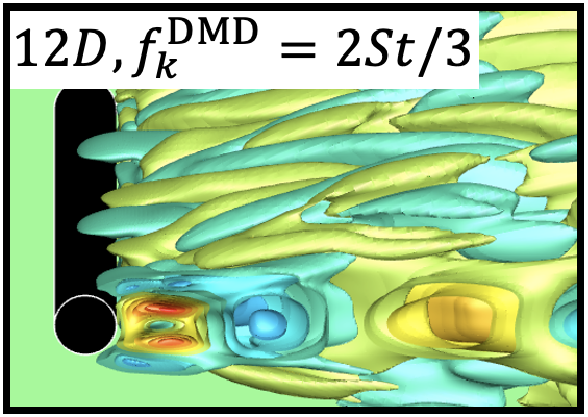}
    %\subcaption{Result at t = 0.5}
  \end{minipage}\\
  \multicolumn{1}{l}{(\textit{c})}  &  \multicolumn{1}{l}{(\textit{d})}\\
  \begin{minipage}[b]{0.5\linewidth}
      %\subcaption{}
          \centering\includegraphics[width=6cm,keepaspectratio]{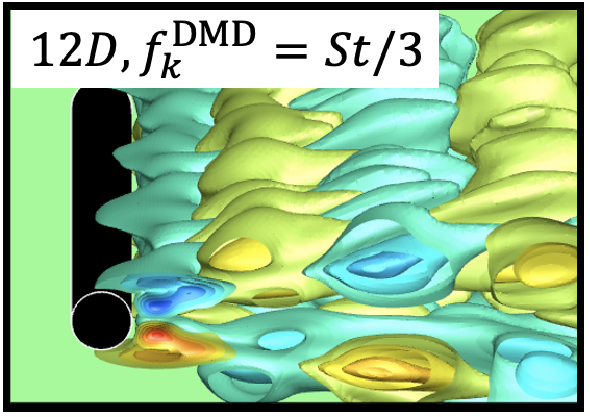}
    %\subcaption{Result at t = 0.25}
  \end{minipage}
  &
  \begin{minipage}[b]{0.5\linewidth}
      %\subcaption{}
          \centering\includegraphics[width=6cm,keepaspectratio]{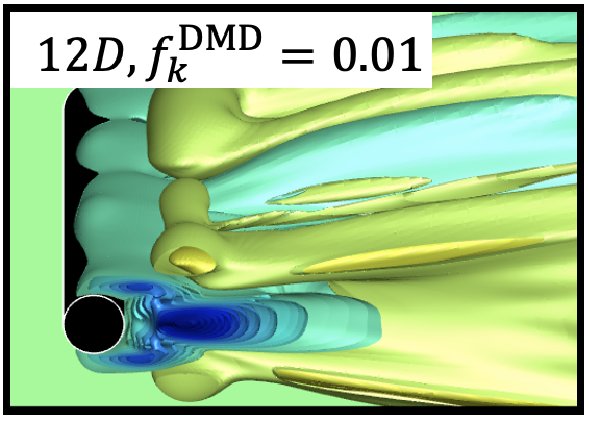}
    %\subcaption{Result at t = 0.5}
  \end{minipage}
   \end{tabular}
    \captionsetup{justification=raggedright,singlelinecheck=false}
  \caption{The same frequency mode as in figure \ref{fig:figure32}, but for $L_z=12D$ case. These distinctive frequency components exist in a fully developed Mode A at frequencies lower than $St$.}
 \label{fig:figure33}
\end{figure}

The low-frequency components shown in figure \ref{fig:figure31}, extracted by the DMD for $L_z=3.4D,$ $3.7D$, and $3.9D$, can be affected by the constraint of spanwise domain size. Before further investigating the extracted low-frequency component, we confirm that similar frequency components exist for a flow field in the sufficiently large domain size. From the distribution of eigenvalues for $L_z=12D$ cases shown in figure \ref{fig:figure3}, eigenvalues close to $f^{\text{DMD}}_k \approx St, 2St/3, St/3,$ and $0.01$ were selected. Figure \ref{fig:figure33} shows the spatial distribution of selected DMD modes. The four DMD modes had a similar distribution to the flow field of small domain sizes, whose numerical constraints were not negligible. Therefore, the existence of the four selected low-frequency components is independent of the effect of the spanwise domain size.

To investigate the shifting process of temporal behavior in more detail, numerical simulations were conducted with finely tuned spanwise domain size, $L_z=3.2D,$ $3.5D,$ $3.6D,$ $3.8D,$ $4.0D,$ $4.2D,$ $4.7D,$ and $5.0D$. For $L_z=3.2D$ and $5.0D$, the two-dimensional flow fields are stable because the growth rate of spanwise velocity was negative. The value of $L_z$ at which the three-dimensional flow was formed coincided with the stable region of the spanwise wavelength of Mode A obtained from the Floquet analysis \citep{Barkley_1996,Rolandi_2023}. 

 \begin{figure}
\begin{tabular}{cc}
\multicolumn{1}{l}{(\textit{a})}  &  \multicolumn{1}{l}{(\textit{b})}\\
  \begin{minipage}[b]{0.8\linewidth}
      %\subcaption{}
          \centering\includegraphics[height=5cm,keepaspectratio]{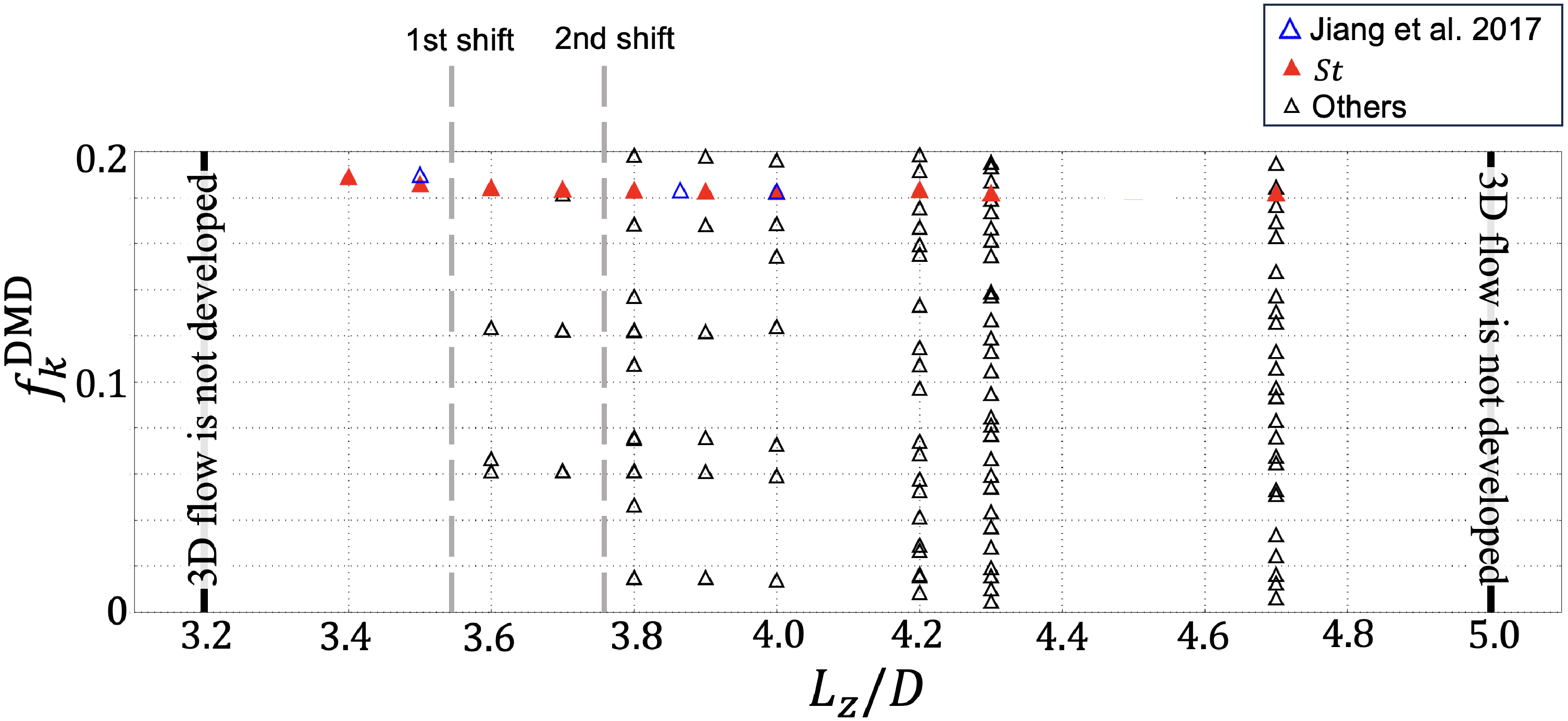}
    %\subcaption{Result at t = 0.5}
  \end{minipage}%\\
 &
   \begin{minipage}[b]{0.2\linewidth}
      %\subcaption{}
          \centering\includegraphics[height=5cm,keepaspectratio]{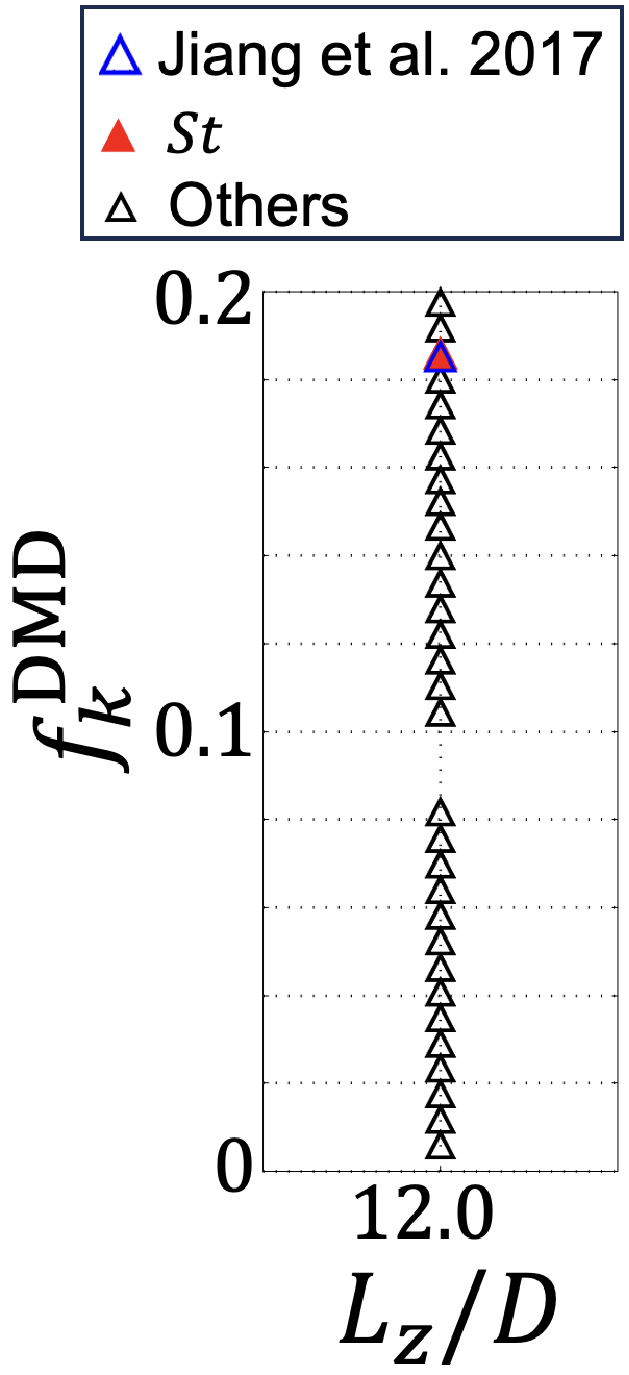}
    %\subcaption{Result at t = 0.25}
  \end{minipage}\hspace{20pt}
  \end{tabular}
    \captionsetup{justification=raggedright,singlelinecheck=false}
  \caption{DMD mode frequencies $f^{\text{DMD}}_k$ in the range of $0$ to $0.2$ at various $L_z$; (\textit{a}) $3.2D\leq L_z \leq 5.0D$, (\textit{b}) $L_z=12D$. Two marked shifts occurred. The second shift is the beginning of the formation of many low-frequency components.}
 \label{fig:figure6}
\end{figure}

DMD was performed on the time-series data of the flow fields with $L_z=3.4D,$ $3.5D,$ $3.6D,$ $3.7D,$ $3.8D,$ $4.0D,$ $4.2D,$ $4.7D,$ and $12D$.  
The $L_z=3.7D$ and $3.8D$ are the transition points where the periodic characteristics of the flow field weaken, and low-frequency components gradually exist from small contribution. To capture the small contribution component for $L_z=3.7D$ and $3.8D$ case, we used the number of ranks $r=194$, which was determined based on the $L_z=12D$ case.
In the other cases, the rank in the SVD approximation was selected based on a cumulative contribution rate exceeding 99.8\%. From the eigenvalue distribution of DMD modes, we extracted stable modes with a growth rate of zero between frequencies $0$ and $0.2$ to focus on the low-frequency component. Figure \ref{fig:figure6} shows the DMD mode frequencies $f^{\text{DMD}}_k$ obtained from the flow fields for each spanwise domain size $L_z$. The frequencies indicated in the red triangle show the $St$ of the wake vortex shedding. $St$ values are in close agreement with the $St$  obtained from the DNS of \citet{Jiang_2017} indicated by the blue triangle.

The first important shift in the appearance of low-frequency components occurred between $3.5D$ and $3.6D$. At this shift, $f^{\text{DMD}}_k = St/3$ and $2St/3$ is emerged. No components other than three frequencies $f^{\text{DMD}}_k = St/3, 2St/3,$ and $St$ appear until the next marked shift occurs between $3.7D$ and $3.8D$. After the shift between $3.7D$ and $3.8D$, a bubble pumping exists. In addition, multiple frequencies appear around $f^{\text{DMD}}_k \approx St, St/3,$ and $2St/3$ simultaneously with the appearance of the bubble pumping. Thereafter, as $L_z$ increases, the number of frequencies appearing around each of $f^{\text{DMD}}_k \approx St, 2St/3,$ $St/3,$ and $0.01$ gradually increases. Therefore, the appearance of multiple frequency components in the flow field begins with the appearance of bubble pumping.

\subsection{BMD-based analysis}\label{BMD}
Previous studies suggest that the low-frequency fluctuations in Mode A are due to the existence of multiple frequencies that are close to $St$ but different from $St$. Following this insight, at the $L_z=3.9D$ case, the sum or difference of the frequencies $St$ and $f^{\text{DMD}}_k \approx 0.01$ was close to the variant frequency around $St$. 
%The emergence process of frequencies in $f^{\text{DMD}}_k \approx St, 2St/3,$ $St/3,$ and $0.01$ suggest the existence of a latent physical mechanism.
 Thus, the sum or difference between two arbitrary frequencies coincides with another frequency. This suggests that the components of these frequencies were related to triadic interactions derived from nonlinearity \citep{phillips_1960}. To detect the interaction relationship in Mode A, we apply BMD to the flow fields at $L_z=3.7D$ and $3.9D$ case.

%\subsection{Interaction with $f=2St/3$ and $St/3$}
BMD was performed on the time-series data for the $L_z=3.7D$ case. The validation of FFT parameters in the BMD algorithm is presented in Appendix B. Figure \ref{fig:figure_37Deigen} shows the eigenvalue distribution obtained from the BMD. Spectral peaks appeared at intervals of $St/3$. This was consistent with the frequency pattern identified in the DMD shown in figure \ref{fig:figure6}. The appearance of the bispectrum peak concludes the triadic interaction between the three frequency components $f \approx St, 2St/3,$ and $St/3$. 
The existence of only integer multiples of $St/3$ in the entire spectral region indicates that one frequency’s integer number frequencies are equal to those of the other frequency. 
The interaction relations for these three frequencies show that the harmonics of the lowest frequency components were equal to the $St$ number. That is the lowest frequency component (third subharmonics), which was exactly ``$1/3$'' of the $St$. The frequency patterns shown in figure \ref{fig:figure6} demonstrate that the number ``$3$'' is universal for at least $3.5D < L_z < 3.9D$. The presence of only doublet waves of the third subharmonic indicates that the overall flow field exhibited periodic behavior at $L_z = 3.7D$. The time variation of the drag coefficient in figure \ref{fig:figure4} (\textit{b}) also shows periodic behavior, which has three different peaks appear repeatedly.  

\begin{figure}
\begin{tabular}{cc}
\multicolumn{1}{l}{(\textit{a})}  &  \multicolumn{1}{l}{(\textit{b})}\\
  \begin{minipage}[b]{0.48\linewidth}
      %\subcaption{}
          \centering\includegraphics[height=12cm,keepaspectratio]{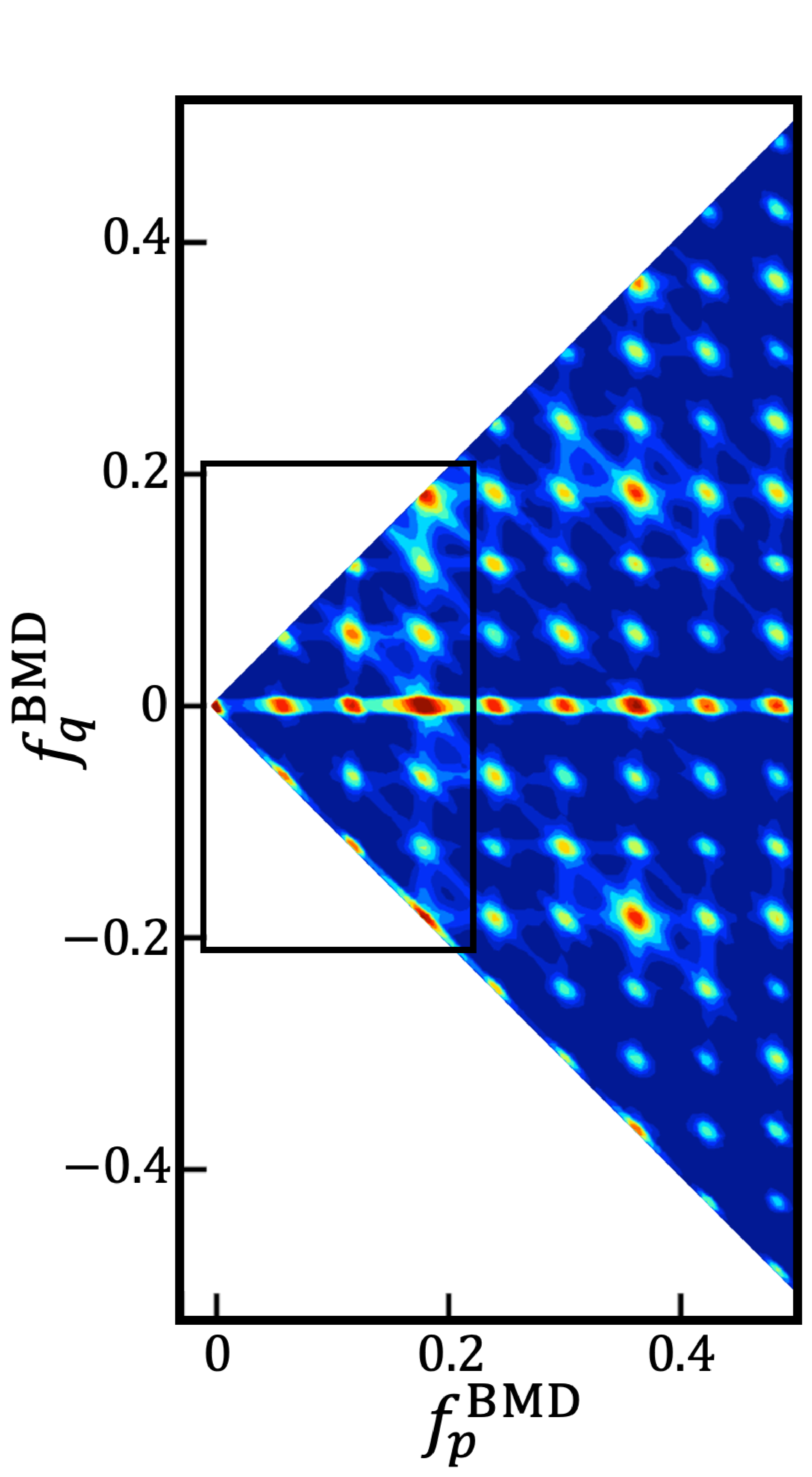}
    %\subcaption{Result at t = 0.25}
  \end{minipage}\hspace{4pt}
  &
  \begin{minipage}[b]{0.48\linewidth}
      %\subcaption{}
          \centering\includegraphics[height=12cm,keepaspectratio]{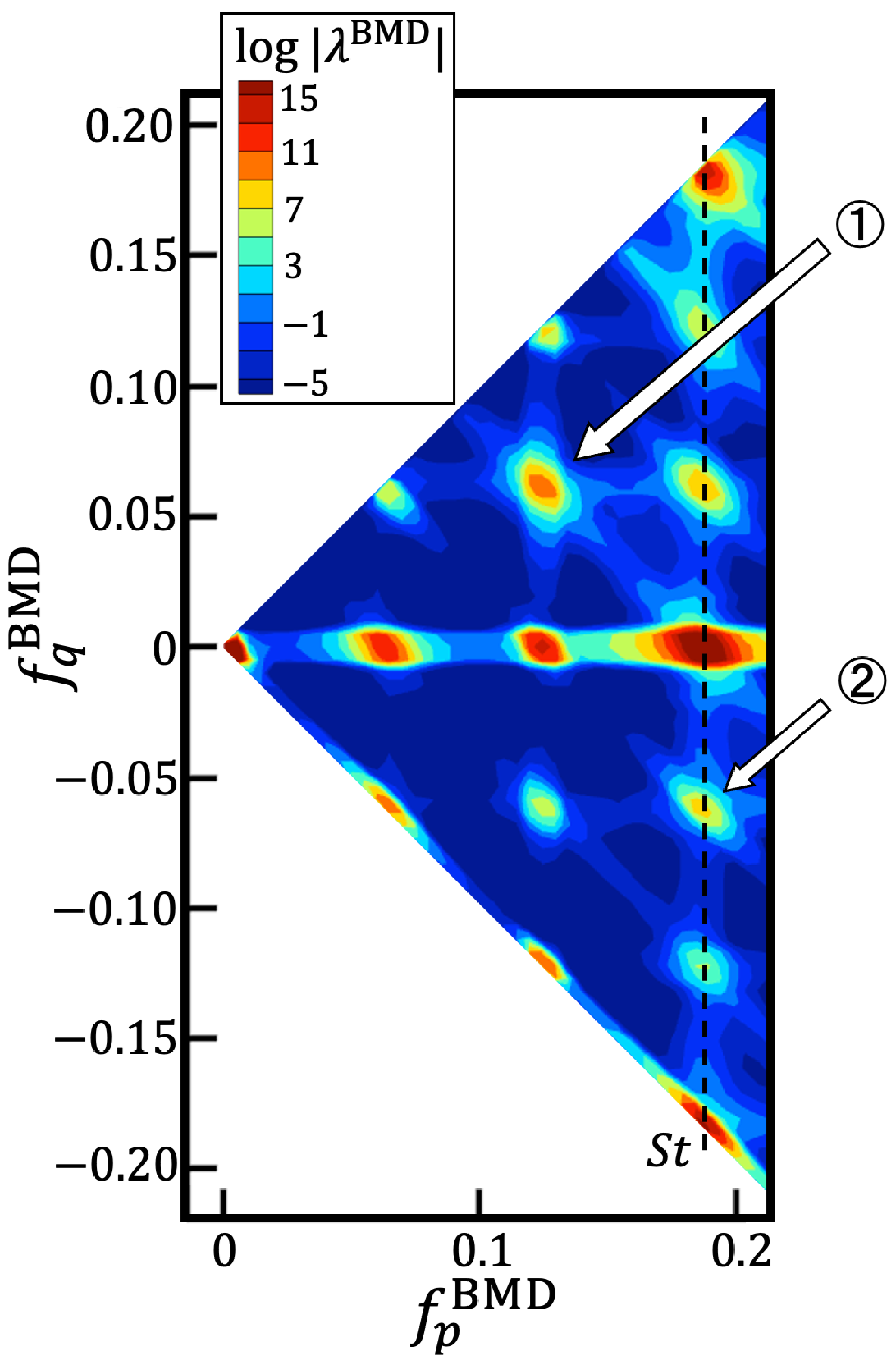}
    %\subcaption{Result at t = 0.5}
  \end{minipage}%\\
  \end{tabular}
    \captionsetup{justification=raggedright,singlelinecheck=false}
  \caption{Eigenvalues corresponding to the two frequencies $f^{\text{BMD}}_p$ and $f^{\text{BMD}}_q$, obtained by the BMD to $L_z=3.7D$: (\textit{a}) wide frequency range, (\textit{b}) close-up view of the frequency range of $0$ to $St$. The absolute values of the eigenvalues are shown in a color map on a logarithmic scale. The interaction is observed only in the harmonics of $St/3$. The mode distributions of \textcircled{1} and \textcircled{2} are shown in figure \ref{fig:figure_37Dmode_1}}
 \label{fig:figure_37Deigen}
\end{figure}

The high peaks are observed at $0$ frequency line $f^{\text{BMD}}_q = 0$, and the points related to $St$ ($\approx 0.18$) such as ($f^{\text{BMD}}_p$, $f^{\text{BMD}}_q$)=($St$, $St$), ($2St$, $St$), ($St$, $-St$), and ($2St$, $-St$). This indicates a large energy cascade into a doublet caused by the Karman vortex. The $f^{\text{BMD}}_p=St$ line indicates that the Karman vortex has a stronger interaction with $St/3$ than with $2St/3$. Therefore, a relationship between $St/3$ and the Karman vortex can be assumed. From a fluid dynamics perspective, a third subharmonic structure can be created from the Karman vortex because the Karman vortex is the most dominant structure caused by the mean flow. %However, the results of the interaction relationship with BMD are not sufficient to reach a conclusion. 

Figure \ref{fig:figure_37Dmode} shows the spatial distributions and interaction relation of the $(f^{\text{BMD}}_p,f^{\text{BMD}}_q) = (2St/3, St/3)$ and $(St, -St/3)$ modes, for which strong interactions were observed in the triad between $f = St/3$, $2St/3$, and $St$. 
 In the interaction $(f^{\text{BMD}}_p,f^{\text{BMD}}_q) = (2St/3, St/3)$, the bispectral mode $\phi_{St/3+2St/3}$ represented a Karman vortex with $f^{\text{BMD}}_{p+q} = St$. For $(f^{\text{BMD}}_p,f^{\text{BMD}}_q) = (St, -St/3)$ interaction, the bispectral mode $\phi_{St-St/3}$ have almost the same spatial distribution as the DMD mode in figure \ref{fig:figure31} (\textit{b}). Because the bispectral mode of a periodic flow coincides with the Fourier mode, and the Fourier mode and the DMD mode at the same frequency exhibit the same spatial distribution \citep{Tu_2013}, the bispectral mode and the DMD mode also exhibit the same spatial distribution. Therefore, the interaction relationship detected by the BMD is comparable to the interaction relationship between the frequency components obtained by the DMD. The interaction map shows a common spatial distribution with strong triad interaction between $f = St/3$, $2St/3$, and $St$. The distribution behind the cylinder in the interaction map clearly shows a common spatial distribution to the $x$-direction in the three frequency components.

\begin{figure}
  \centering\includegraphics[width=14cm,keepaspectratio]{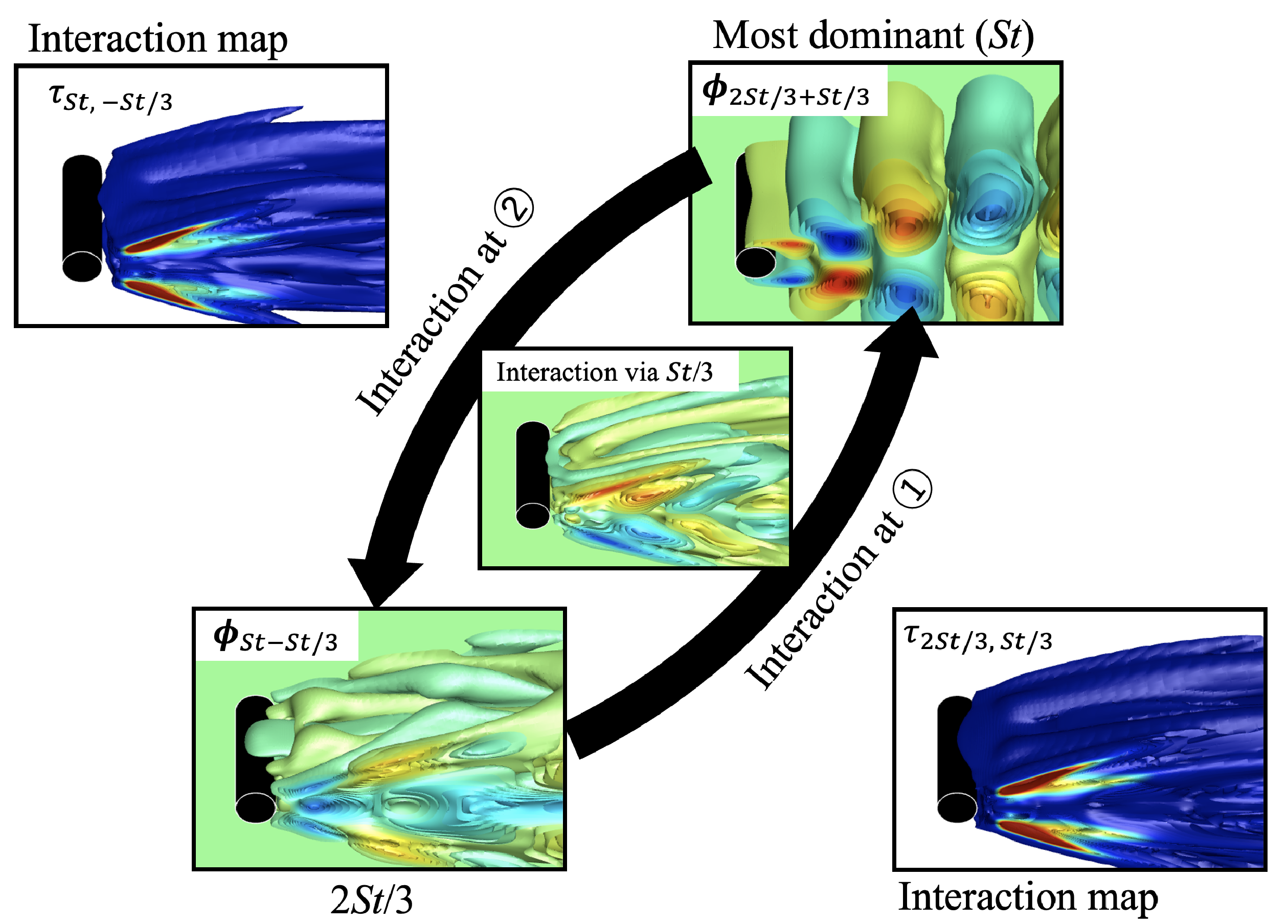}
    \captionsetup{justification=raggedright,singlelinecheck=false}
  \caption{Isosurface of bispectral mode, interaction map obtained from $L_z=3.7D$ case. All isosurface represents $x$-direction velocity component. BMD indicates the triad relationship of the three frequency components $f = St/3$, $2St/3$, and $St$.}
 \label{fig:figure_37Dmode_1}
\end{figure}

%\subsection{Interaction with bubble pumping}
\begin{figure}
\begin{tabular}{cc}
\multicolumn{1}{l}{(\textit{a})}  &  \multicolumn{1}{l}{(\textit{b})}\\
  \begin{minipage}[b]{0.45\linewidth}
      %\subcaption{}
          \centering\includegraphics[height=10cm,keepaspectratio]{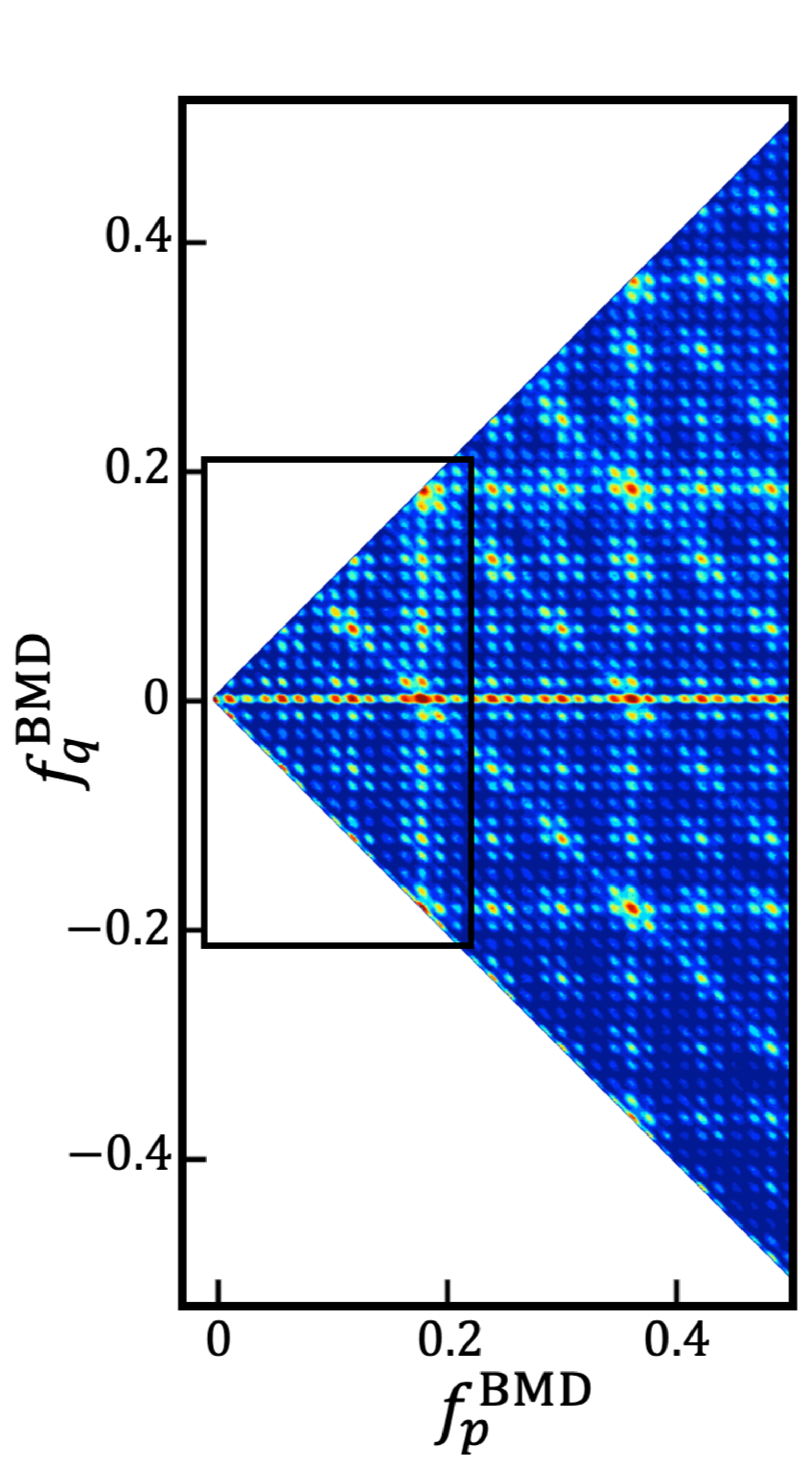}
    %\subcaption{Result at t = 0.25}
  \end{minipage}
  &
  \begin{minipage}[b]{0.54\linewidth}
      %\subcaption{}
          \centering\includegraphics[height=10cm,keepaspectratio]{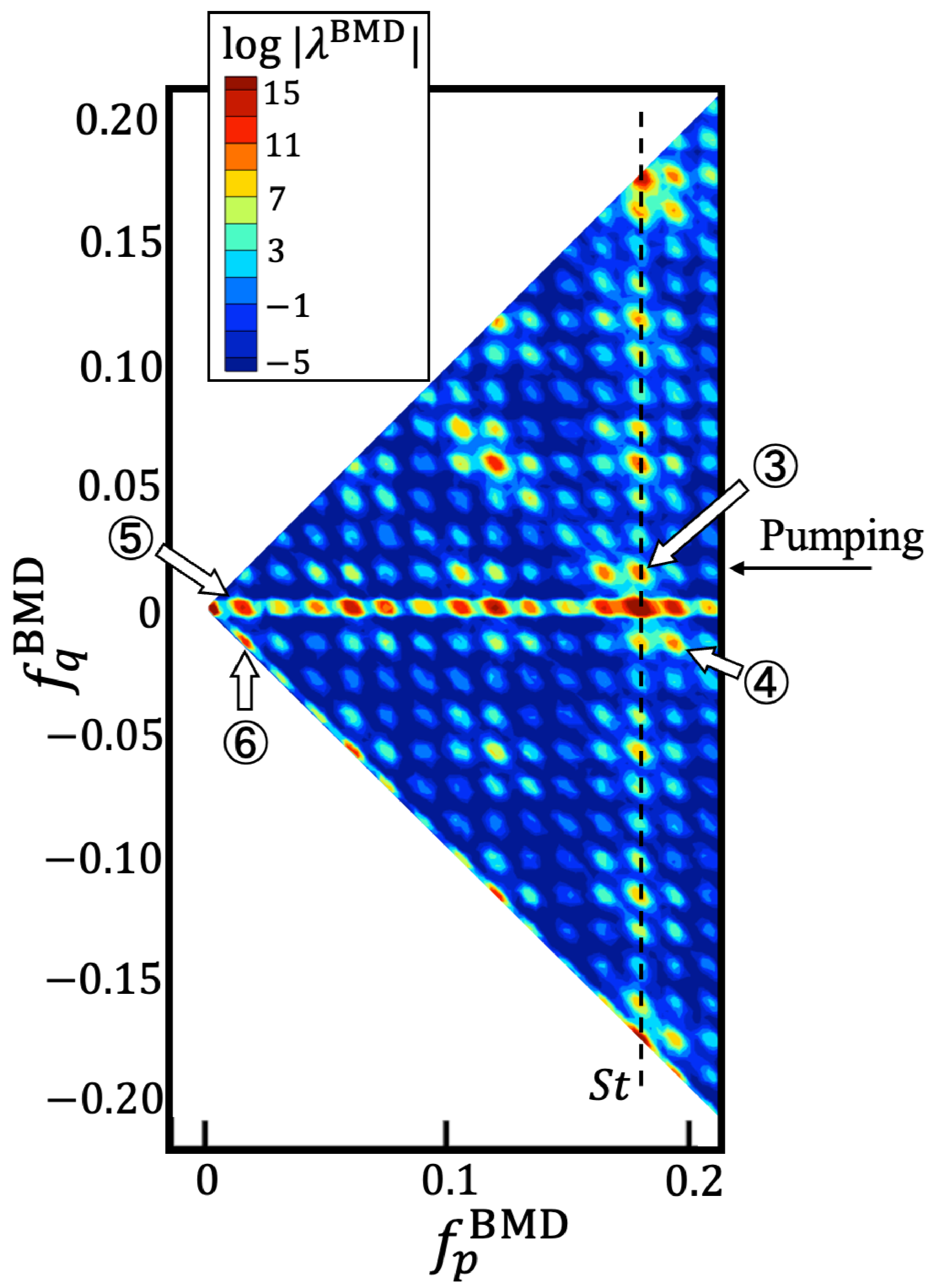}
    %\subcaption{Result at t = 0.5}
  \end{minipage}%\\
  \end{tabular}
    \captionsetup{justification=raggedright,singlelinecheck=false}
  \caption{Eigenvalues corresponding to the two frequencies $f^{\text{BMD}}_p$ and $f^{\text{BMD}}_q$, obtained by the BMD to $L_z=3.9D$: (\textit{a}) wide frequency range, (\textit{b}) close-up view of the frequency range of $0$ to $St$.The absolute values of the eigenvalues are shown in a color map on a logarithmic scale. With the appearance of a bubble pumping, a sub-peak appears around the main peak frequency. The mode distributions of \textcircled{3}, \textcircled{4}, \textcircled{5}, and \textcircled{6}, the interaction of the Karman vortex and the bubble pumping, are shown in figure \ref{fig:figure_39Dmode_1} and figure \ref{fig:figure_39Dmode_2}.}
 \label{fig:figure_39Deigen}
\end{figure}
We performed BMD on the $L_z=3.9D$ case to identify the interaction involved in the bubble pumping. Figure \ref{fig:figure_39Deigen} shows the distribution of the eigenvalues obtained from the BMD. Numerous peaks were observed compared with the eigenvalue distribution of the BMD at $L_z=3.7D$, which had no frequency classified as bubble pumping. The primary peaks were observed at $f = St/3, 2St/3, St$, same for $L_z=3.7D$ case. However, in this case, subpeaks existed around the largest peak at $f = St/3, 2St/3, St$. The existence of subpeaks was different from the BMD eigenvalue distribution in the $L_z=3.7D$ case. The line with $f^{\text{BMD}}_q \approx 0.01$ has a peak at the point $(f^{\text{BMD}}_p,f^{\text{BMD}}_q)  \approx (0.01, St/3),$ $(0.01, 2St/3),$ and $(0.01, St)$ which is interaction points between the primary frequency and bubble pumping. As a result of these interactions, the bispectrum of frequency $f^{\text{BMD}}_{p+q}$ is reinforced, and the neighboring peak around the primary peak $f^{\text{BMD}}_p$, thus $f^{\text{BMD}}_p+0.01$ becomes strength. Thus, these sub-peaks around the primary peaks indicated the interaction between the bubble pumping and the three frequencies.

In the $L_z=3.9D$ case, the sub-peaks distant from the primary peak, such as the peak at $f^{\text{BMD}} =(St/3+2 \times 0.01), (St/3-2 \times 0.01),$ and $(St -2 \times 0.01)$ were relatively small. That is, the interaction between the sub-peak frequency and bubble pumping is relatively small in this case. However, figure \ref{fig:figure6} shows that, as spanwise domain size $L_z$ increased, the number of frequency components obtained by the DMD increased. This means frequency components that had weak peaks gradually became strong peaks and could be captured by the DMD. This indicates that the interaction with bubble pumping becomes strong as the spanwise boundary size increases. Notably, the $L_z=12D$ in figure \ref{fig:figure6} (\textit{a}) showed that the interval between frequencies obtained by the DMD was approximately constant. The value of these intervals was approximately equal to the frequency of the bubble pumping. This means many frequencies appear in the $L_z=12D$, resulting in the interaction with the bubble pumping. Thus, in a flow field with a sufficiently large spanwise domain and unconstrained by domain size, non-linear interaction with bubble pumping increases the number of frequency components in the overall flow field.

\begin{figure}
  \centering\includegraphics[width=14cm,keepaspectratio]{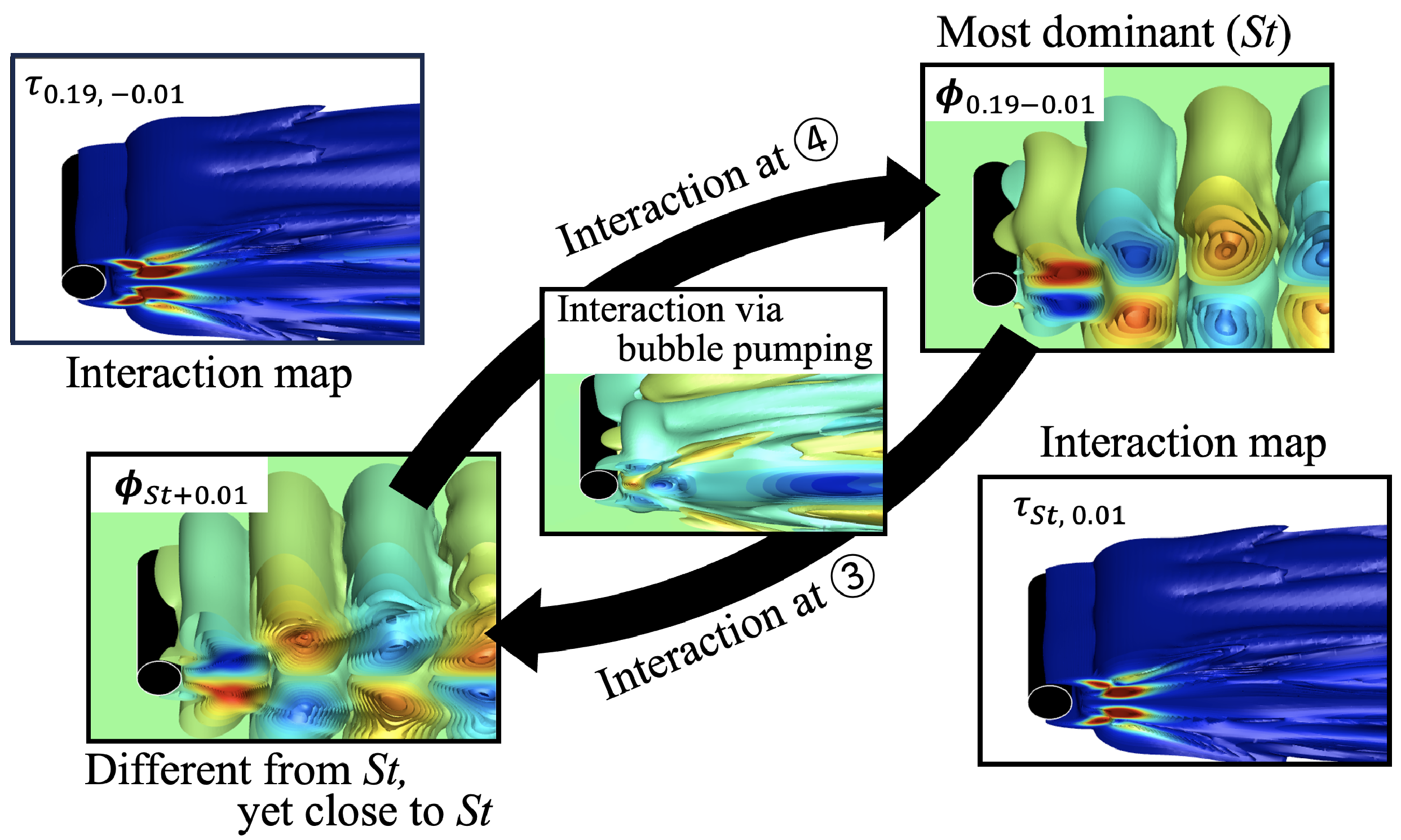}
    \captionsetup{justification=raggedright,singlelinecheck=false}
  \caption{Isosurface of bispectral modes, interaction maps at $(f^{\text{BMD}}_p,f^{\text{BMD}}_q) \approx (St, 0.01)$ and $(0.19, -0.01)$ obtained from $L_z=3.9D$ case. All isosurface is $x$-direction velocity component.}
 \label{fig:figure_39Dmode_1}
\end{figure}

We focused on the mode distributions in the interaction points in which large peaks were observed at figure \ref{fig:figure_39Deigen}. Figure \ref{fig:figure_39Dmode_1} shows the spatial distributions of $(f^{\text{BMD}}_p,f^{\text{BMD}}_q) \approx (St, 0.01)$ and $(0.19, -0.01)$ modes and these interaction relation. These are the interactions via bubble pumping between the primary peak and its neighboring sub-peaks. The bispectral modes $\phi_{0.19-0.01}$ and $\phi_{St+0.01}$ represent both wake Karman vortices at different frequency. Hence, the presence of Karman vortices at different frequencies is associated with low-frequency fluctuations, which supports the assertions of \citet{Henderson,Jiang_2017}. These interactions are characterized by bubble pumping. The interaction map shows strong interaction near the $x$-axis of the cylinder wake. Moreover, the interaction is strong only at the position near the cylinder. This indicates that the frequency of the Kármán vortex in the wake is governed by the shedding timing at the cylinder.

\begin{figure}
  \centering\includegraphics[width=14cm,keepaspectratio]{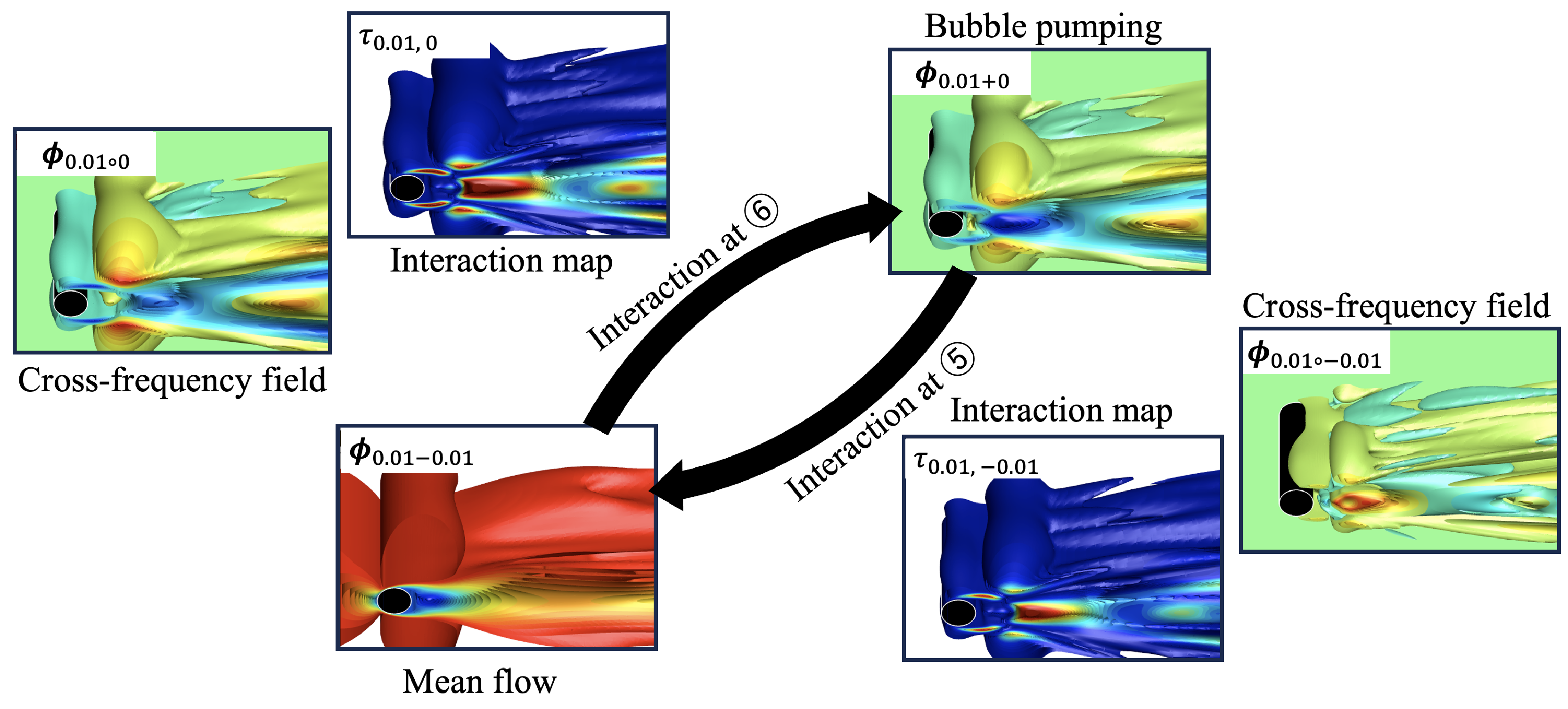}
  \captionsetup{justification=raggedright,singlelinecheck=false}
  \caption{Isosurface of cross-frequency fields, bispectral modes, interaction maps at $(f^{\text{BMD}}_p,f^{\text{BMD}}_q) \approx (0.01, -0.01)$ and $(0, 0.01)$ obtained from $L_z=3.9D$ case. All isosurface is $x$-direction velocity component.}
 \label{fig:figure_39Dmode_2}
\end{figure}

Figure \ref{fig:figure_39Dmode_2} shows the spatial distributions of $(f^{\text{BMD}}_p,f^{\text{BMD}}_q) \approx (0.01, 0)$ and $(0.01, -0.01)$ modes and these interaction relation. The interaction related to $0$-frequency component (mean flow) and the bubble pumping, $(f^{\text{BMD}}_p,f^{\text{BMD}}_q) \approx (0.01, 0)$, $(0.01, -0.01)$ has a symmetric structure about the $x$-axis. The cross-frequency fields $\phi_{0 \circ 0.01}$ and $\phi_{0.01 \circ -0.01}$ are distributed around the wake recirculation region in the $0$-frequency mode, which is the bispectral mode $\phi_{0.01-0.01}$. The interaction map also shows a strong interaction around the recirculation region. These distributions imply correlations between the vortices formed in the recirculation region and the bubble pumping.

\section{Emergence of bubble pumping and its effect on flow fields}\label{summalize}
The appearance of the bubble pumping can be considered the beginning of the complexity of the flow field. Therefore, it is important to further investigate bubble pumping.
\subsection{Harmonic nature between the lowest-frequency mode and Karman vortex}
The BMD implies that the $f=St/3$-mode is the third-subharmonic to the Karman vortex. On the contrary, the harmonic relationship between the frequencies of the bubble pumping and the Karman vortex requires further investigation. Table \ref{table_harmonic} lists the most dominant Karman vortex frequency $St$, the lowest frequency $g$ obtained from the DMD, and $St/g$. The value of $St/g$ below the decimal point at $L_z=3.6D$ and $3.7D$, where no bubble pumping is observed, is about $10$ times smaller than those at the other $L_z$ values where a bubble pumping exists. Although the integer part of the $St/g$ varied with $L_z$, the difference below the decimal point was substantial. This clearly indicated that the $St$ did not belong to the harmonics of the bubble pumping. %The appearance of low-frequency components that deviate from the Karman vortex subharmonic is 

\begin{table}
 \centering
  \begin{tabular}{ccccccccc}
   & $3.6D$ & $3.7D$ & $3.8D$ & $3.9D$ & $4.0D$ & $4.2D$ & $4.3D$ & $4.7D$ \\
   $St$ & $0.18421$ & $0.18354$ & $0.18326$ & $0.18281$ & $0.18245$ & $0.18338$ & $0.18181$ & $0.18182$ \\
   $g$ & $0.060950$ & $0.061013$ & $0.014819$ & $0.014871$ & $0.013884$ & $0.008401$ & $0.004626$ & $0.006116$ \\
   $St/g$ & $3.022$ & $3.008$ & $12.366$ & $12.293$ & $13.141$ & $21.828$ & $39.303$ & $29.729$ \\
  \end{tabular}
     \captionsetup{justification=raggedright,singlelinecheck=false}
  \caption{Harmonic nature of various $L_z$ indicated by the most dominant Karman vortex frequency $St$, the lowest frequency $g$, and $St/g$. }
 \label{table_harmonic}
\end{table}

\subsection{Relationship with 0-frequency mode}
The self-interaction of the bubble pumping, which is the interaction point at \textcircled{6} in figure \ref{fig:figure_39Deigen}, is stronger than the interaction with the Karman vortex, which is the interaction point at \textcircled{3} and \textcircled{4}. Thus, the bubble pumping is strongly related to the $0$-frequency component compared to the Karman vortex. 
\begin{figure}
    \centering\includegraphics[width=12cm,keepaspectratio]{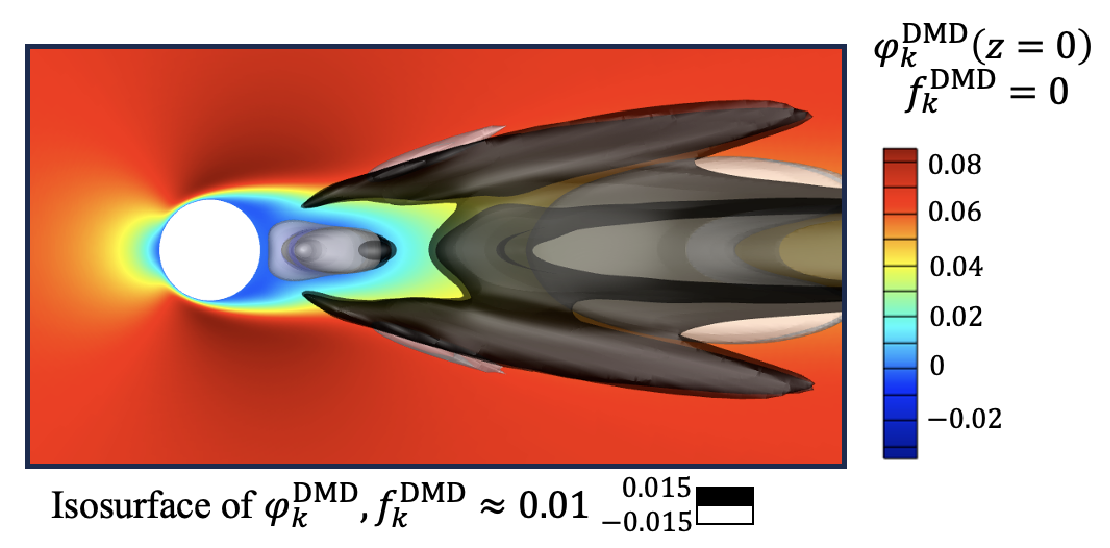}
    \captionsetup{justification=raggedright,singlelinecheck=false}
  \caption{Overlay plot of cross-section at $z=0$ for the real part of the $x$-direction velocity component of the DMD mode with $f_k^{\text{DMD}}=0$ and an isosurface for the DMD mode with $f_k^{\text{DMD}} \approx 0.01$ obtained from $L_z=3.9D$ case.}
 \label{fig:figure_pumping_overlay}
\end{figure}
Figure \ref{fig:figure_pumping_overlay} shows a cross-section at $z=0$ for the DMD mode with $f_k^{\text{DMD}}=0$ and an isosurface for the DMD mode with $f_k^{\text{DMD}} \approx 0.01$ obtained from the $L_z=3.9D$ case. The symmetric structure of the bubble pumping to the $x$-axis was similar to the $0$-frequency component. Furthermore, the DMD mode at $f_k^{\text{DMD}} \approx 0.01$ was distributed along the wake recirculation region of the $0$-frequency component. The interaction maps and spatial distribution of the cross-frequency fields of the BMD, as shown in figure \ref{fig:figure_39Dmode_2}, also supported these distribution relations. \citet{Yokota_2025} also experimentally pointed out the relationship between the recirculation region and the bubble pumping. Based on these results, the bubble pumping is related to the distribution of the recirculation region formed behind the cylinder.

\begin{figure}
  \centering\includegraphics[width=10cm,keepaspectratio]{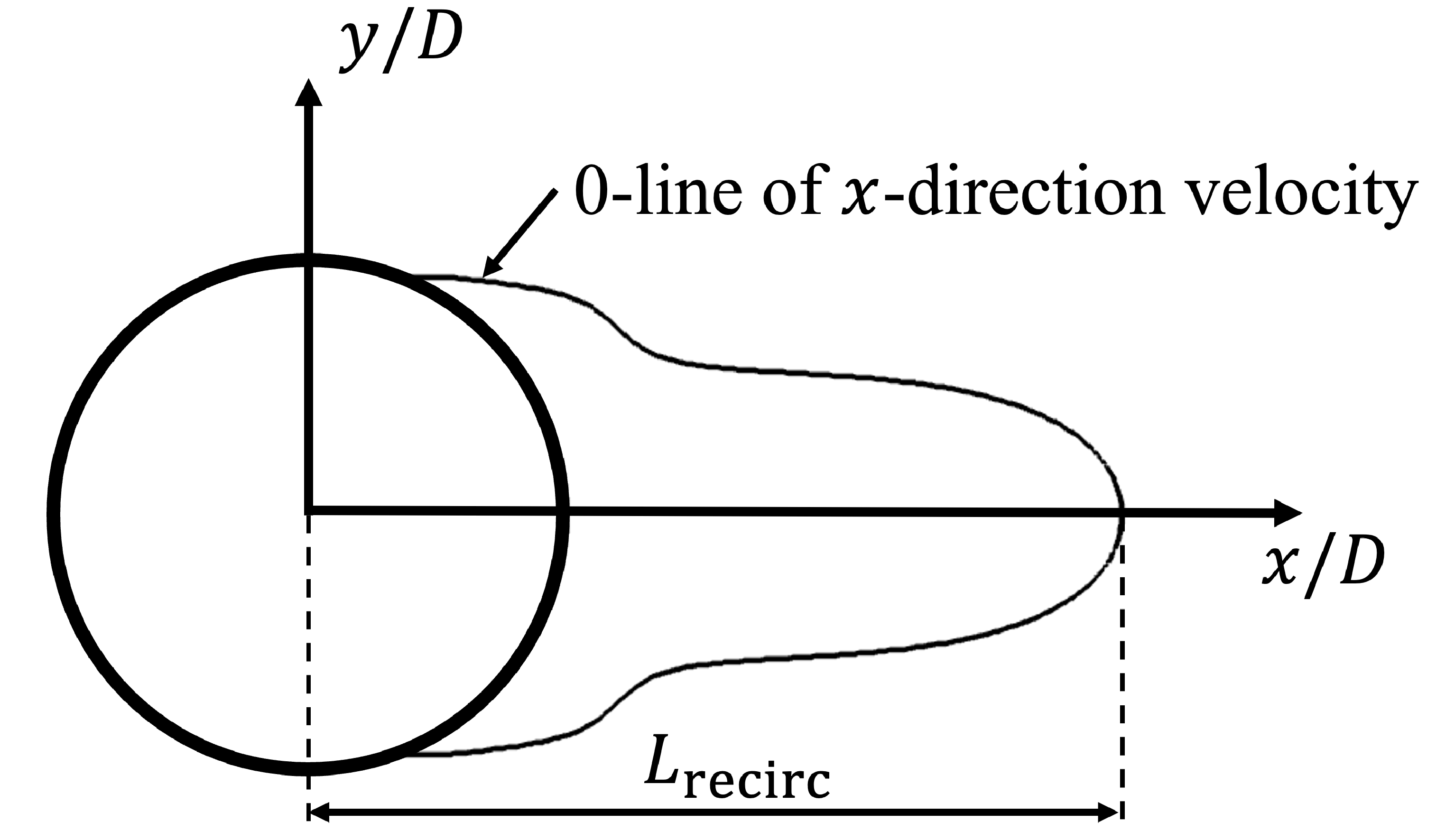}
    \captionsetup{justification=raggedright,singlelinecheck=false}
  \caption{Definition of $L_\text{recirc}$.}
 \label{fig:Lrecirc}
\end{figure}

\begin{figure}
    \centering\includegraphics[width=14cm,keepaspectratio]{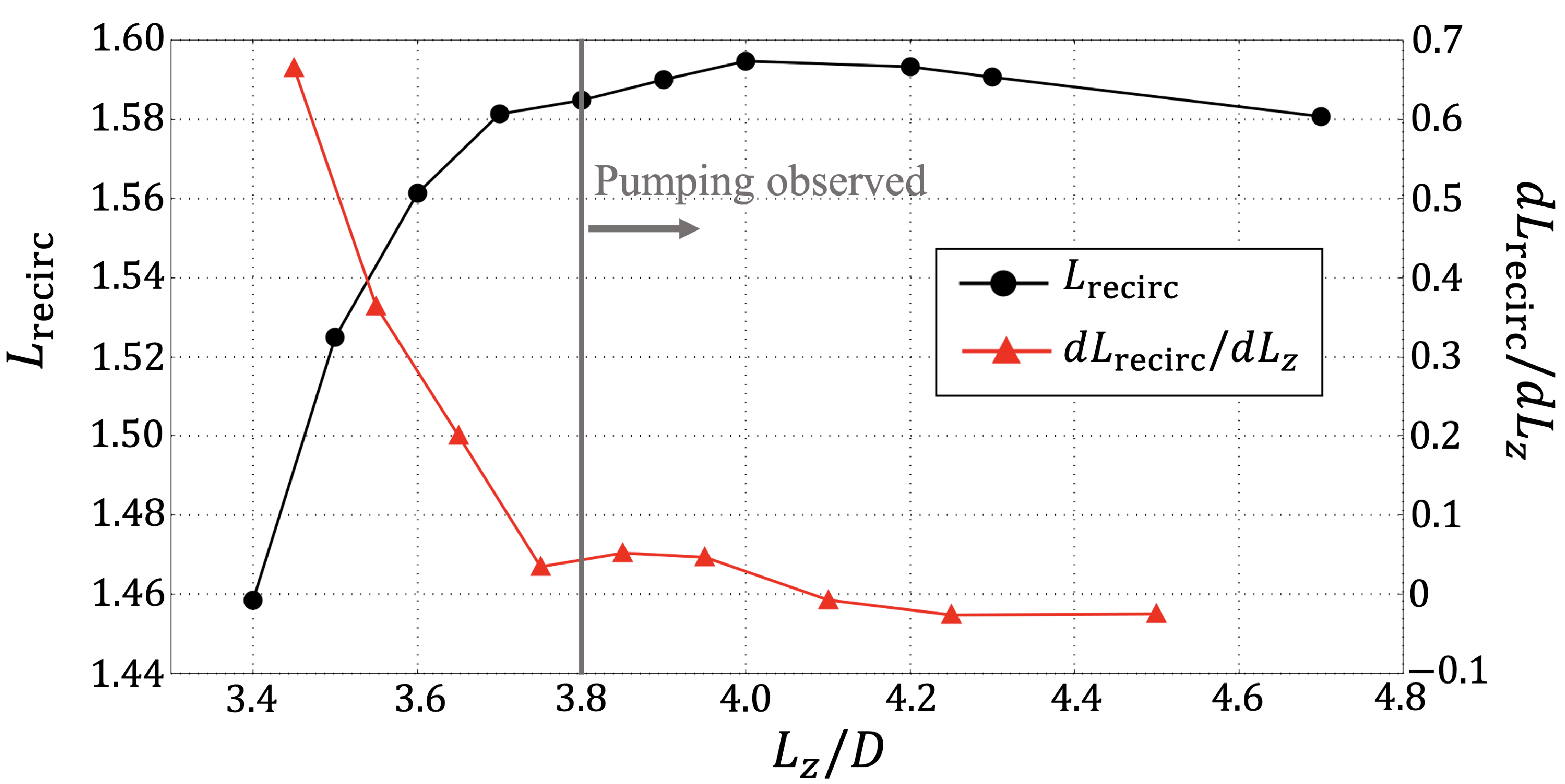}
    \captionsetup{justification=raggedright,singlelinecheck=false}
  \caption{The length of the recirculation region $L_\text{recirc}$ and gradient of $L_\text{recirc}$ to the $L_z$. The increase in length of the recirculation region satuates with the appearance of the bubble pumping at $3.7D < L_z < 3.8D$.}
 \label{fig:figure_pumpingLz}
\end{figure}

Numerous studies on the recirculation region of cylindrical wakes \citep{Fornberg_1980,mypaper_steady} have shown that the length in the main flow direction is a parameter that characterizes the recirculation region. Figure \ref{fig:Lrecirc} shows a conceptual diagram of the recirculation region formed in the cylinder wake. In this study, the length of the recirculation region $L_\text{recirc}$ was the normalized $x$-coordinate at which the time and spanwise averages of the velocity in the main flow direction became $0$ except at the nonslip surface of the cylinder.
 Figure \ref{fig:figure_pumpingLz} shows the values of the recirculation region $L_\text{recirc}$ for various $L_z$ and the gradient of $L_\text{recirc}$ with respect to $L_z$. The gradient for $L_z$ was computed from the two reference points $L_z=L_1$ and $L_2$ using the central difference with the gradient $L_z=\frac{L_1+L_2}{2}$ as follows:
\begin{equation}
\frac{dL_\text{recirc}}{dL_z} \bigg|_{L_z = \frac{L_1 + L_2}{2}} 
= \frac{L_\text{recirc}(L_1) - L_\text{recirc}(L_2)}{L_1 - L_2}.
\label{recircgrad}
\end{equation}
The grey lines in figure \ref{fig:figure_pumpingLz} emphasize that the bubble pumping was observed at $L_z\geq3.8D$ from the frequency distribution obtained with the DMD shown in figure \ref{fig:figure6}.

For $L_z<3.8D$, $L_\text{recirc}$ was extended with increasing $L_z$. In other words, at $L_z<3.8D$, the size of the recirculation region expanded with an increase in the spanwise wavelength, which was equal to $L_z$. The end of this expansion of $L_\text{recirc}$ was clear when viewed in terms of the gradient of $L_\text{recirc}$ with respect to $L_z$. Therefore, the size of the recirculation region became constant with the appearance of a bubble pumping at $L_z=3.8D$.

Since the DMD mode with $f_k^{\text{DMD}}=0.01$ is distributed around the recirculation region, it can be more controversial to reconstruct the flow field by these two modes. Reconstruction using the selected DMD mode is computed as follows:
 \begin{equation}
  \begin{split}
\boldsymbol{u}'(\boldsymbol{x},t_j)=[\boldsymbol{\varphi}^{\text{DMD}}_{k_1}(\boldsymbol{x}),\boldsymbol{\varphi}^{\text{DMD}}_{k_2}(\boldsymbol{x}), \cdots \boldsymbol{\varphi}^{\text{DMD}}_{k_s}(\boldsymbol{x})][b_{k_1}(t_j), b_{k_2}(t_j), \cdots b_{k_s}(t_j)]^T\\
  \label{Aapproximation}
  \end{split}
\end{equation}
%\text{where} b_k(t_j)={\boldsymbol{\varphi}^{\text{DMD}}_k}^{+}\boldsymbol{u}(\boldsymbol{x},t_j),
where subscript $k_j$ indicates the index of selected DMD mode, $b_k(t_j)$ is the coupling coefficient for the DMD mode $\boldsymbol{\varphi}^{\text{DMD}}_k$ in the $t=t_j$, and directly computed from the flow snapshot as follows:
 \begin{equation}
  \begin{split}
[b_{k_1}(t_j), b_{k_2}(t_j), \cdots b_{k_s}(t_j)]=[\boldsymbol{\varphi}^{\text{DMD}}_{k_1}(\boldsymbol{x}),\boldsymbol{\varphi}^{\text{DMD}}_{k_2}(\boldsymbol{x}), \cdots \boldsymbol{\varphi}^{\text{DMD}}_{k_s}(\boldsymbol{x})]^{\dagger}\boldsymbol{u}(\boldsymbol{x},t_j).\\
  \label{calcoefficient}
  \end{split}
\end{equation}
Here, the Moore–Penrose pseudoinverse matrix is computed by preconditioning the QR decomposition. 

\begin{figure}
    \centering\includegraphics[width=14cm,keepaspectratio]{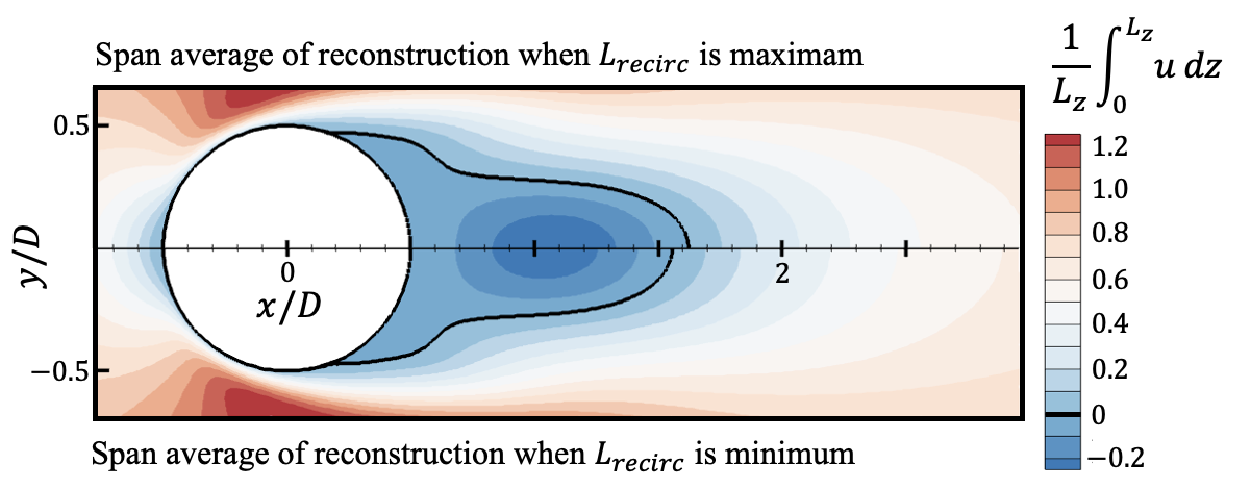}
    \captionsetup{justification=raggedright,singlelinecheck=false}
  \caption{Spatial distribution of the spanwise average of the reconstructed flow field using the DMD mode with $f^{\text{DMD}}_k=0$ and $0.01$ for the time when the recirculation region is the maximum (top) and minimum (bottom). The bubble pumping represents the growth and compression of the recirculation region.}
 \label{fig:figure_pumping}
\end{figure}

At the $L_z=3.9D$ case, the coupling coefficients $b_k(t_j)$ for the DMD modes representing the bubble pumping and $0$-frequency field were computed, and the flow fields were reconstructed by the $0$-frequency field and DMD mode at $f_k^{\text{DMD}} \approx 0.01$ and $-0.01$. Time-series data for the reconstructed flow field were averaged over the spanwise direction. From the span averaged flow fields, the time variation of $L_\text{recirc}$ was obtained. Figure \ref{fig:figure_pumping} shows the instantaneous fields when $L_\text{recirc}$ is at its maximum and minimum. In the two instantaneous fields, the difference in $L_\text{recirc}$ was approximately $0.1$. The temporal fluctuating component of the reconstructed flow field was limited to the structure originating from the bubble pumping because the mean field had no temporal fluctuating component. Thus, the bubble pumping represent the expansion and compression of the recirculation region. 

Back to the relationship between $L_z$ and $L_\text{recirc}$ in figure \ref{fig:figure_pumpingLz}, as $L_z$ increases, $L_\text{recirc}$ expands. As $L_z$ increases in size beyond $3.8D$, althogh the time-average of $L_\text{recirc}$ remains constant, but the maximum lengths that the recirculation region takes could continue to increase with $L_z$. Because the most stable spanwise wavelength for Mode A is around $L_z \approx 3.8D$ \citep{Henderson,cylinder4,Jiang_2017}, the constant value of $L_\text{recirc}$ at $L_z>3.8D$ is considered to be determined by the stability of the flow field. That is, bubble pumping is a low-frequency oscillation of recirculation around a stable recirculation region. In terms of relationship with spanwise domain size, the increase in $L_z$ enables an expansion of the recirculation region beyond the stable $L_\text{recirc}$, and bubble pumping becomes a permissible.

\subsection{Interaction between Karman vortex and bubble pumping}

\begin{figure}
\begin{tabular}{c}
\multicolumn{1}{l}{(\textit{a})}\\
  \begin{minipage}[b]{1.0\linewidth}
      %\subcaption{}
          \centering\includegraphics[width=14cm,keepaspectratio]{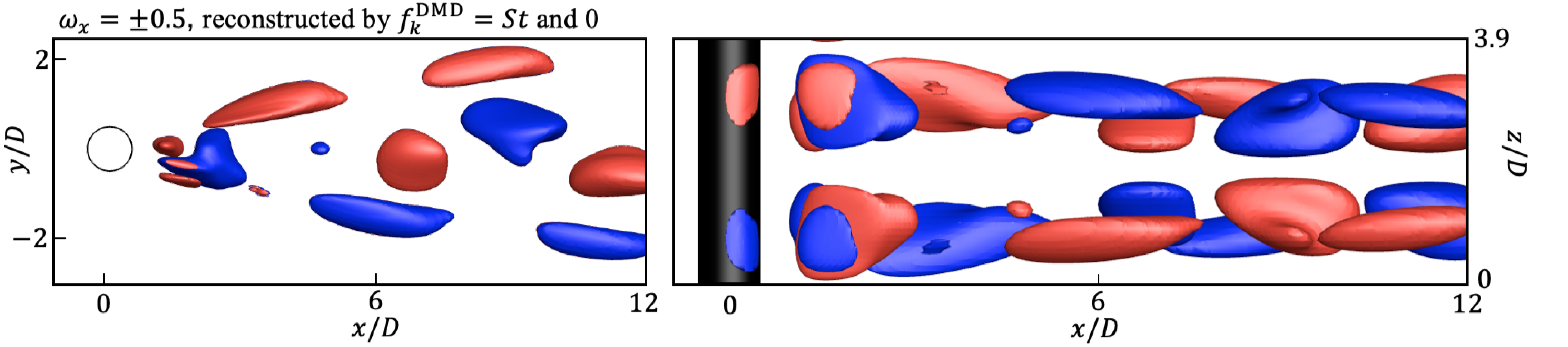}
    %\subcaption{Result at t = 0.25}
  \end{minipage}\\
  
  \multicolumn{1}{l}{(\textit{b})}\\
  \begin{minipage}[b]{1.0\linewidth}
      %\subcaption{}
          \centering\includegraphics[width=14cm,keepaspectratio]{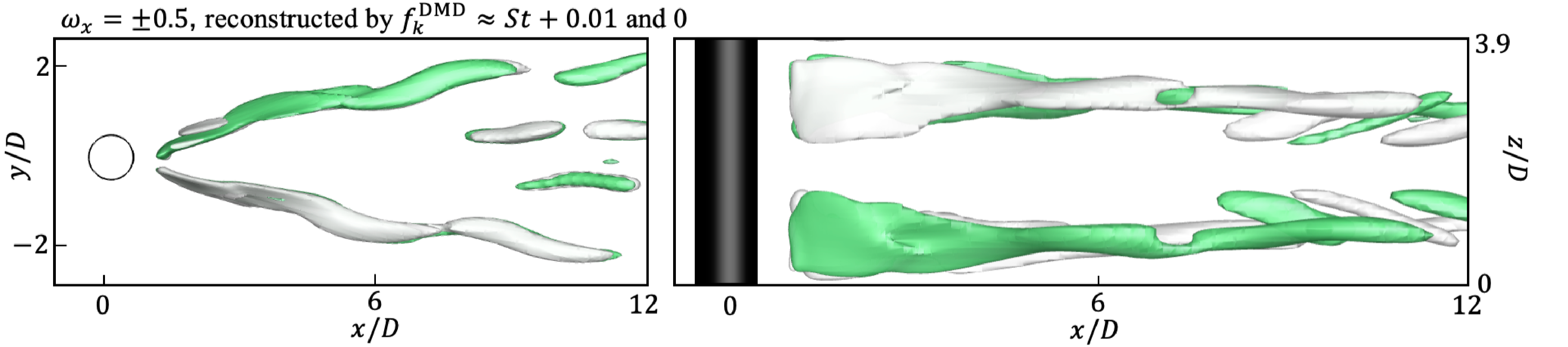}
    %\subcaption{Result at t = 0.5}
  \end{minipage}\\
  
    \multicolumn{1}{l}{(\textit{c})}\\
  \begin{minipage}[b]{1.0\linewidth}
      %\subcaption{}
          \centering\includegraphics[width=14cm,keepaspectratio]{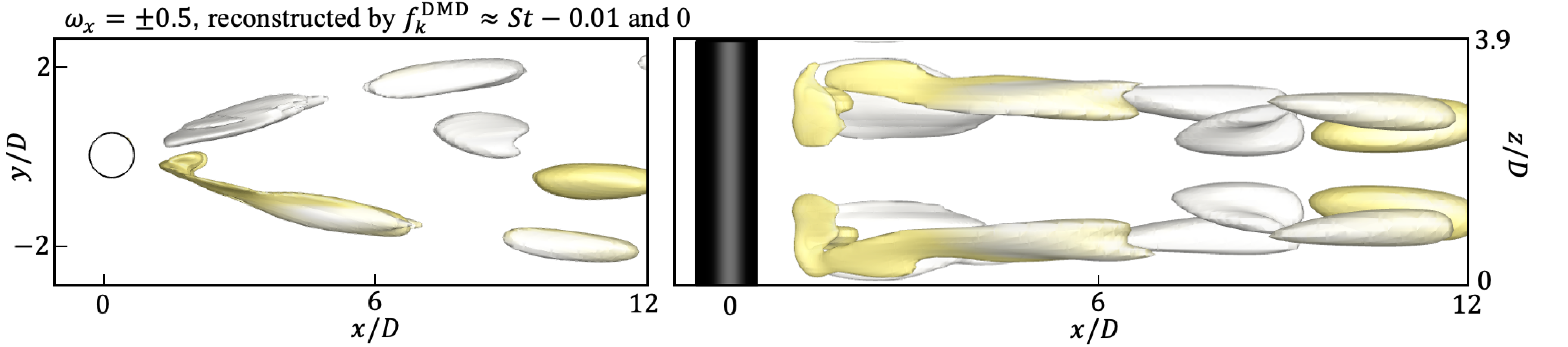}
    %\subcaption{Result at t = 0.5}
  \end{minipage}\\
  
    \multicolumn{1}{l}{(\textit{d})}\\
  \begin{minipage}[b]{1.0\linewidth}
      %\subcaption{}
          \centering\includegraphics[width=14cm,keepaspectratio]{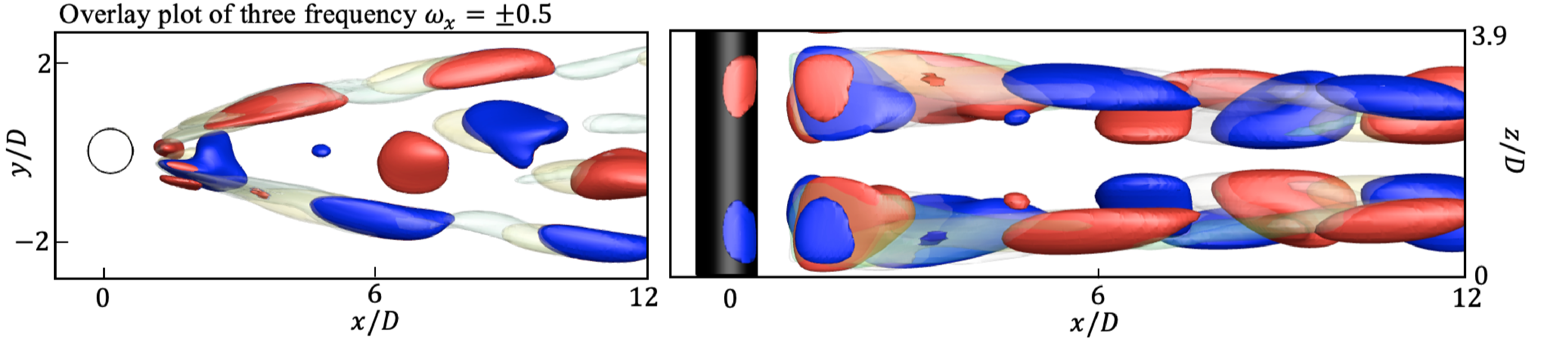}
    %\subcaption{Result at t = 0.5}
  \end{minipage}\\
  \end{tabular}

  \captionsetup{justification=raggedright,singlelinecheck=false}
  \caption{Isosurfaces of $x$-direction vorticity $\omega_x$ computed from the flow field reconstructed from the DMD mode: (\textit{a}) $f^{\text{DMD}}_k=0$ and $St$, (\textit{b})  $f^{\text{DMD}}_k=0$ and $St+0.01$, and (\textit{c}) $f^{\text{DMD}}_k=0$ and $St+0.01$, (\textit{d}) is an overlay plot of three vorticity isosurfaces. Vortices of $f^{\text{DMD}}_k \approx St+0.01$ and $St-0.01$ are distributed around a vortex of $f^{\text{DMD}}_k=St$. Owing to the interaction with the bubble pumping, fluctuations appear in the vortex of $f^{\text{DMD}}_k=St$.}
 \label{fig:figure_pumping_vol}
\end{figure}

We reconstructed the vortex structure at frequencies $St$, $St+0.01$, and $St-0.01$ using the DMD modes of $0$-frequency and the corresponding frequency at $L_z=3.9D$ case. The vortices of $f^{\text{DMD}}_k \approx St+0.01$ and $St-0.01$ are considered to have appeared due to the interaction between the bubble pumping and the Karman vortex.
A fully developed Mode A forms a distinctive periodic structure in the spanwise direction, thus the isosurface of the $x$-direction vorticity characterazes the wake Karman vortex indicated by \citet{cylinder4,Jiang_2017,modeA}.
 From the reconstructed flow field, the vorticity in the $x$-direction was calculated as follows:
\begin{equation}
\omega_x = \frac{D}{U_\infty} \left( \frac{\partial w'}{\partial y} - \frac{\partial v'}{\partial z} \right).
\label{eqvolx}
\end{equation}
where $w'$ and $v'$ represent the reconstructed spanwise and transverse velocity component, respectively.
Figure \ref{fig:figure_pumping_vol} shows the instantaneous fields of the isosurface in $\omega_x$ at the following three frequencies: $f^{\text{DMD}}_k = St$, $f^{\text{DMD}}_k \approx  St+0.01$, and $St-0.01$. Three different frequency vortices are distributed at the same spatial location. Hence, these vortices are not independent, but form one vortex street in the interaction with each other.  Because $St$ is the most dominant vortex shedding frequency, primary vortex is represented by $f^{\text{DMD}}_k = St$. 

Figure \ref{fig:figure_pumping_vol} (\textit{d}) shows overlay plots for isosurfaces of the three-frequency vorticities. The overlay plot of the three vortices shows that the vortices of the two frequencies other then $St$ are distributed around the primary vortex $f=St$. This seems to be the result of the most dominant Karman vortex saptialy fluctuating, producing the vortex of different frequencies. This scenario of the primary Karman vortex fluctuating spatially is due to the bubble pumping in the wake, since the fluctuations appear with the interaction of the frequency components representing the bubble pumping. The same fluctuations occur in all principal frequency components, since the BMD results show interaction of bubble pumping with third-subharmonics and harmonics other than the Karman vortex. Hence, with the emergence of bubble pumping, the significant coherent structure flucuates spatially, leading to the emergence of various frequency components.

\section{Conclusions}\label{conclude}
This study focuses on apearance of low-frequency fluctuation in the Mode A and its effect on the flow fields. The existence of low-frequency component in the Mode A at a $Re=200$ is confirmed when the spanwise domain size $L_z$ in the numerical simulation is changed. The low-frequency components were identified in the numerical simulation under $L_z$ where a pair of Mode A structures existed in the computational domain ($3.2D < L_z < 5.0D$).  Applying the DMD to numerical results with various $L_z$ revealed the frequencies and corresponding spatial structure inherent in the shifting process of the temporal behavior related to apearance of low-frequency component. A marked shift occurred between $L_z=3.5D$ and $3.6D$ and $L_z=3.7D$ and $3.8D$. In the former shift, oscillations at frequencies of $St/3$ and $2St/3$ appeared, and in the latter, a frequency of about $0.01$ appeared. 

With the aid of BMD analysis for the case of $L_z=3.7D$, interactions exist between $f=St/3$, $2St/3$, and $St$, and the flow field consists of a $St/3$ harmonic. Existance of only harmonics component maintains the periodic behavior of flow fields. When $L_z=3.9D$, the bubble pumping interacted with the other frequency components such as $f=St/3, 2St/3,$ and $St$. Moreover, the appearance of bubble pumping also disrupts the harmonic nature with respect to the lowest-frequency component. Due to the decay of these harmonic properties, interactions with bubble pumping generated numerous frequency peaks in the spectrum.

The BMD interaction map between bubble pumping and mean flow shows a strong interaction around the recirculation region, and reveals a strong relationship between the mean field and bubble pumping. From the span and time averaged flow fields, the length of the recirculation region in $0$-frequency mode remained constant with the appearance of a bubble pumping when the spanwise domain size gradually increases. 
The reconstruction of the flow field from the DMD mode corresponding to the $0$-frequency mode and bubble pumping showed that the bubble pumping represented the expansion and compression of the recirculation region. Hence, bubble pumping is a phenomenon in which the recirculation region fluctuates around a stable length. 

%The interaction map between Karman vortex and bubble pumping shows strong interaction near the cylinder. 
From the  interaction map between Karman vortex and bubble pumping, and reconstruction using DMD modes, it was discovered that the nonlinear interaction between the most dominant Karman vortex and the bubble pumping caused spatial fluctuations in the dominant Karman vortex. It is due to fluctuations originating from bubble pumping in the recirculation region near the cylinder. Thus, the interaction between the bubble pumping and the dominant coherent structure near the cylinder causes fluctuations, leading to an increase in the number of frequency components in the overall flow field.

The constraint from spanwise domain sizes revealed several significant frequency components hidden within the complex Mode A. However, the mechanism underlying the appearance of the $f=St/3$ and $2St/3$ modes requires further investigation. In addition, a reasonable explanation for the appearance of the third subharmonic of the Karman vortex is required in future studies. Future studies on Mode B and A at different $Re$ values are also expected to reveal new physical mechanisms. 

\appendix
\section{Grid and time dependance}\label{appen1}
\begin{figure}
    \centering\includegraphics[keepaspectratio, scale=0.3]{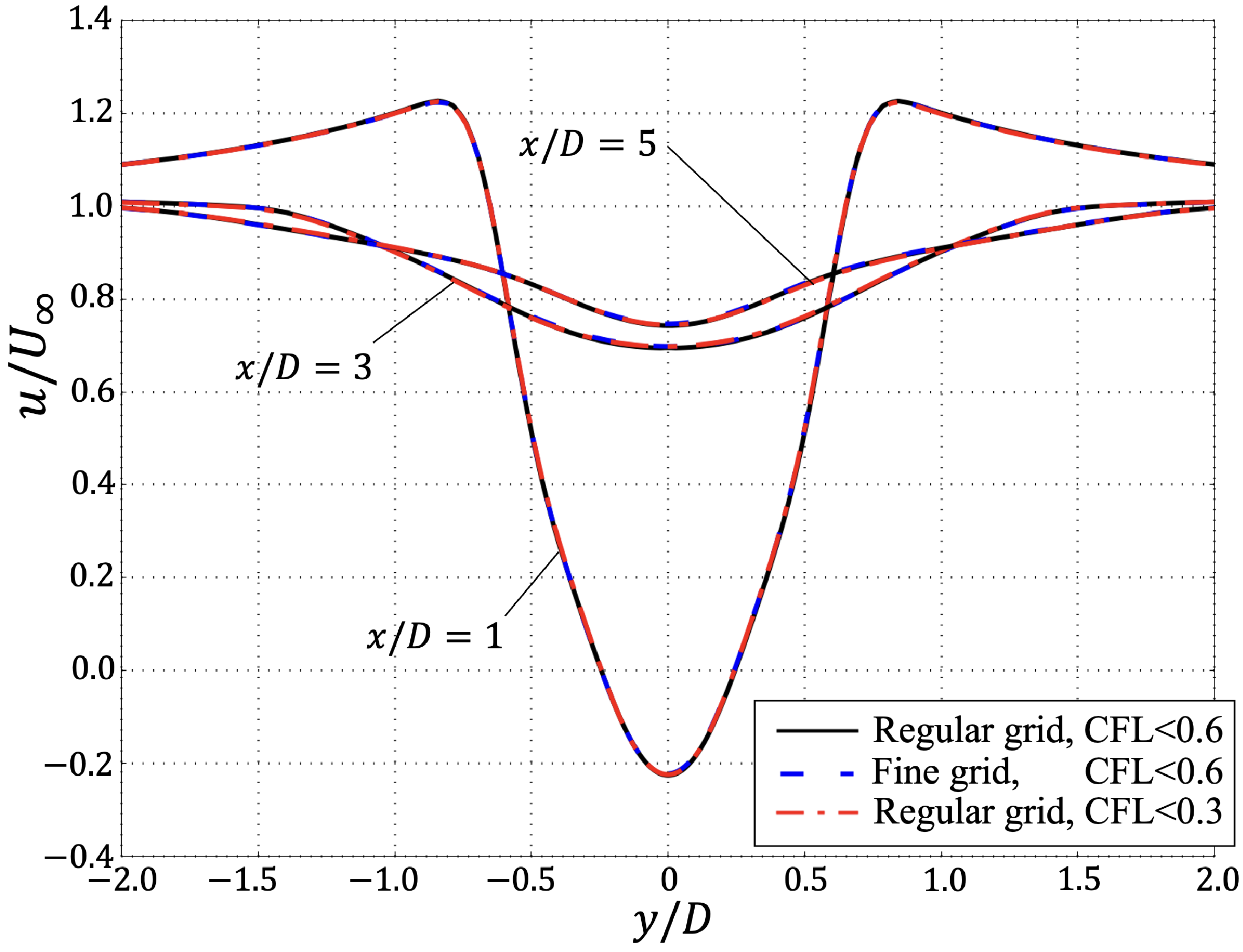}
    \captionsetup{justification=raggedright,singlelinecheck=false}
  \caption{Comparison with average fields of streamwise velocity at $Re=300$.}
 \label{fig:figurea}
\end{figure}

The computational mesh and CFL numbers were selected based on a convergence study. A convergence test was performed by comparing the results obtained on the regular grid and CFL number with those obtained on a finer grid and smaller CFL number. The number of cells for the fine grid was $320$ in the wall-normal direction and $660$ in the wall-parallel direction. The CFL number for the small CFL cases was set to 0.3 or less. This dependence was tested in a flow field of $Re=300$ to ensure that the dissipation effects derived from the time-step size and grid width were not dominant at $Re=200$. Figure \ref{fig:figurea} shows a comparison of the $x$-directional velocities averaged over time and spanwise directions for $x/D=1$, $x/D=3$, and $x/D=5$. The numerical dissipation, which depends on the grid width, was not dominant because the average results for a fine grid were not different from those for a regular grid. When CFL was halved, the average time remained constant. Therefore, the computational result of CFL$<0.6$ for the regular grids was reasonable, and all $L_z$ cases were computed under CFL$<0.6$ and had the same grid width as regular grids.

\section{Convergence study for BMD}\label{appen3}
We addressed the selection of parameters for the FFT dataset within the BMD algorithm. Table \ref{table2} lists the estimation parameters of the spectrum for $L_z=3.7D$ and $3.9D$. The time-step size of the datasets was $0.1$, and the Nyquist frequency was $5$ in all cases. $N_\text{FFT}$ was determined based on the sampling frequency requirement of the spectrum. $L_z=3.9D$ required a higher frequency resolution than $L_z=3.7D$ to capture the lowest frequency structures belonging to bubble pumping.

The number of blocks $N_\text{blk}$ was verified to ensure that the spectrum converged. The convergence was evaluated by integrating the power spectral density over the entire domain.
\begin{table}
 \centering
  \begin{tabular}{ccccc}
   $\,\,\,L_z\,\,\,$ & $\,\,\,N_{\text{FFT}}\,\,\,$  & $\,\,\,N_{\text{ovlp}}\,\,\,$ & $\,\,\,\Delta t\,\,\,$ & $\,\,\,N_{\text{blk}}\,\,\,$\\[3pt]
   $3.7D$ & $2048$ & $1536$  & $0.1$ & $7$\\
   $3.9D$ & $4096$ & $3072$  & $0.1$ & $9$\\
  \end{tabular}
   \caption{Spectrum estimation parameters in the FFT for each case.}
 \label{table_FFTpara}
  \label{table2}
\end{table}
 \begin{equation}
  \begin{split}
\lambda^{\text{FFT}}(f)=\sum^{N_\text{blk}}_{k=1}\sum^{N}_{j=1} \{ \hat{\boldsymbol{u}}^{k*}(\boldsymbol{x}_j,f) \circ \hat{\boldsymbol{u}}^k(\boldsymbol{x}_j,f) \}.
  \label{FFTspec}
  \end{split}
\end{equation}
%\lambda^{\text{FFT}}(\textit{f})=\sum^{N_\text{blk}}_{k=1} \int \{ \hat{\boldsymbol{u}}^{k*}(\boldsymbol{x},f) \circ \hat{\boldsymbol{u}}^k(\boldsymbol{x},f) \} d\boldsymbol{x}.
\begin{figure}
\begin{tabular}{c}
\multicolumn{1}{l}{(\textit{a})}\\
  \begin{minipage}[b]{0.95\linewidth}
      %\subcaption{}
          \centering\includegraphics[width=12cm, keepaspectratio]{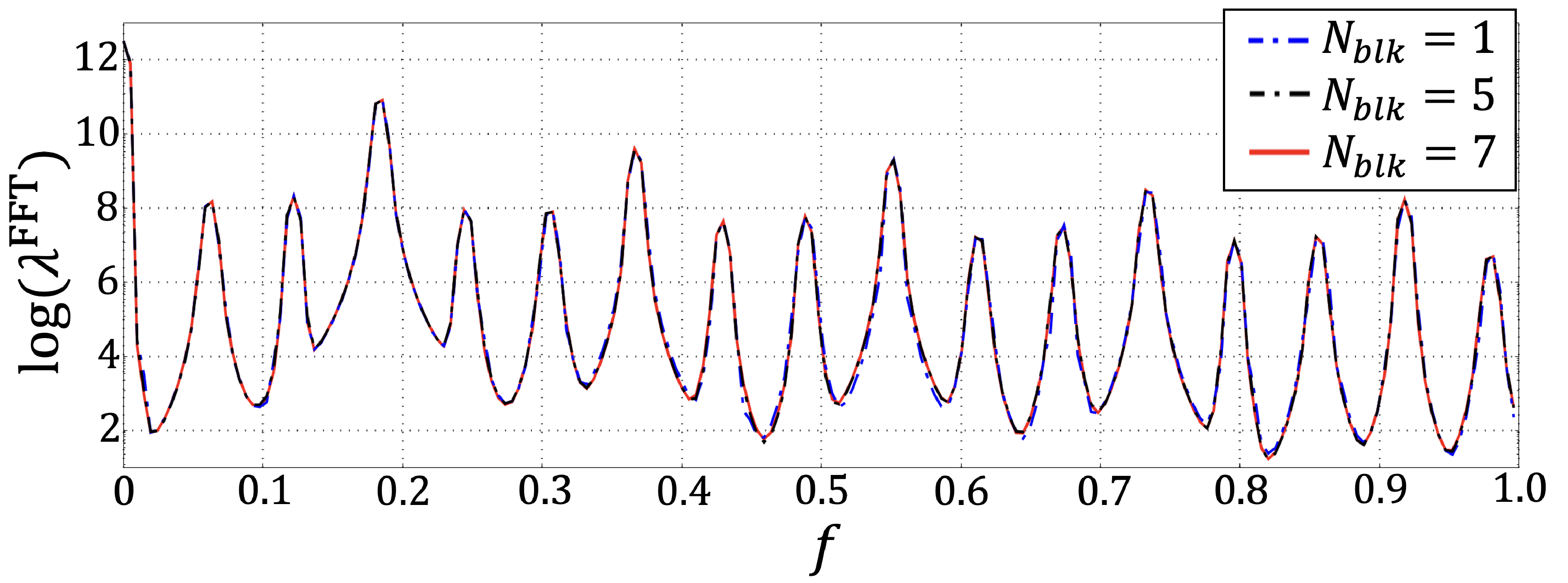}
    %\subcaption{Result at t = 0.25}
  \end{minipage}\\
\multicolumn{1}{l}{(\textit{b})}\\
  \begin{minipage}[b]{0.95\linewidth}
      %\subcaption{}
         \centering\includegraphics[width=12cm, keepaspectratio]{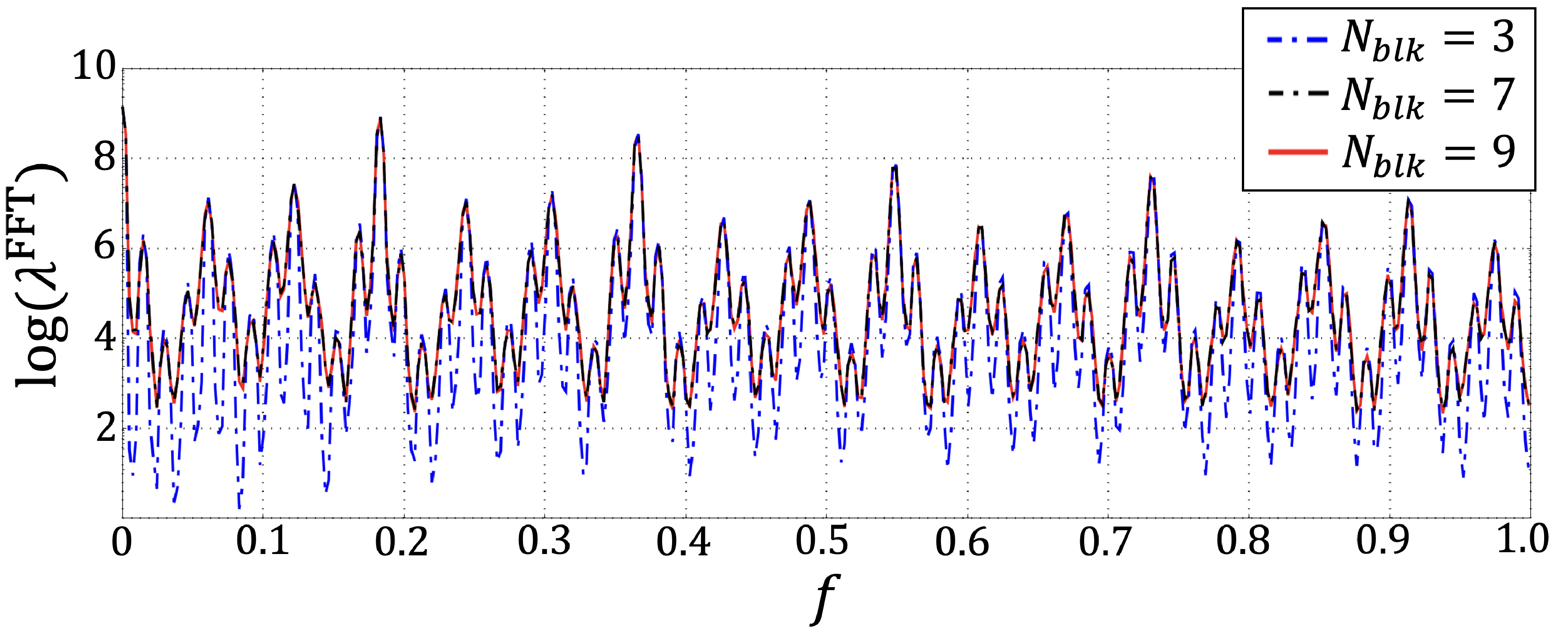}
    %\subcaption{Result at t = 0.5}
  \end{minipage}\vspace{-5pt}\\
  \end{tabular}
    \captionsetup{justification=raggedright,singlelinecheck=false}
  \caption{Frequency spectrum for increasing the number of blocks in the FFT; (\textit{a}) $L_z=3.7D$, (\textit{b}) $L_z=3.9D$. In this paper, we adopted $N_{\text{blk}}=7$ for $L_z=3.7D$ and $N_{\text{blk}}=9$ for $L_z=3.9D$.}
 \label{fig:figureb}
\end{figure}
Figure \ref{fig:figureb} shows the value of $\lambda^{\text{FFT}}(\textit{f})$ as the number of blocks increased. The Hanning window was used for spectral estimation. For $L_z=3.7D$, $\lambda^{\text{FFT}}(\textit{f})$ with $N_{\text{blk}}=5$ and $7$ blocks is consistent. The same applies for $L_z=3.9D$, where $N_{\text{blk}}=7$ and $9$. Therefore, the choice of block numbers $N_{\text{blk}}=7$ and $9$ for each FFT at $L_z=3.7D$ and $3.9D$, respectively, is reasonable.

\backsection[Acknowledgements]{
We thank Prof. Taku Nonomura's pointing to the homogeneous nature of the spanwise direction.
We gratefully acknowledge Dr. Yuya Ohmichi and Yasuhito Okano for providing valuable knowledge about bubble pumpings and Hiroki Sakamoto for advice on oblique instability. We thank Yuta Iwatani for the valuable discussions on nonlinear interactions and BMD. }

\backsection[Funding]{The numerical simulations were performed on the supercomputer systems “AFI-NITY” and “AFI-NITY II” at the Advanced Fluid Information Research Center, Institute of Fluid Science, Tohoku University, and JAXA Supercomputer System Generation 3 (JSS3). 
This study was partially supported by a Sasakawa Scientific Research Grant from the Japan Science Society. This study was partially supported by JST SPRING, Grant Number JPMJSP2114, Japan. }

\backsection[Declaration of interests]{The authors report no conflict of interest.}

%\backsection[Supplementary data]{\label{SupMat}Supplementary material and movies are available at \\https://doi.org/10.1017/jfm.2019...}

%\backsection[Data availability statement]{The data that support the findings of this study are openly available in [repository name] at http://doi.org/[doi], reference number [reference number]. See JFM's \href{https://www.cambridge.org/core/journals/journal-of-fluid-mechanics/information/journal-policies/research-transparency}{research transparency policy} for more information}

\backsection[Author ORCIDs]{Authors may include the ORCID identifers as follows.  Y. Nakamura, https://orcid.org/0009-0008-4118-1078; S. Sato, https://orcid.org/0000-0002-9979-0051; N. Ohnishi, https://orcid.org/0000-0001-5895-0381}

%\backsection[Author contributions]{Authors may include details of the contributions made by each author to the manuscript'}

%\bibliographystyle{jfm}
%\bibliography{jfm}
%Use of the above commands will create a bibliography using the .bib file. Shown below is a bibliography built from individual items.

%\bibliographystyle{plain}  % または適切なスタイルを使用
%\bibliography{references}

\bibliographystyle{jfm}

\end{document}